\documentclass[12pt]{article}
\usepackage{amssymb,graphicx,epsfig} 
\def\ltap{\raisebox{-.4ex}{\rlap{$\sim$}} \raisebox{.4ex}{$<$}} 
\def\gtap{\raisebox{-.4ex}{\rlap{$\sim$}} \raisebox{.4ex}{$>$}} 
\def\etal{{\em et al.}}
\def\issue(#1,#2,#3){{\bf #1}, #2 (#3)} 

\def\APP(#1,#2,#3){{\it Acta Phys.\ Polon.} \ \issue({\bf #1},#2,#3)}
\def\ARNPS(#1,#2,#3){{\it Ann.\ Rev.\ Nucl.\ Part.\ Sci.} \ \issue({\bf #1},#2,#3)}
\def\CPC(#1,#2,#3){{\it Comp.\ Phys.\ Comm.} \ \issue({\bf #1},#2,#3)}
\def\CIP(#1,#2,#3){{\it Comput.\ Phys.} \ \issue({\bf #1},#2,#3)}
\def\EPJ(#1,#2,#3){{\it Eur.\ Phys.\ J.} \ \issue({\bf #1},#2,#3)}
\def\EPJD(#1,#2,#3){Eur.\ Phys.\ J. Direct\ C \ \issue({\bf #1},#2,#3)}
\def\IEEETNS(#1,#2,#3){{\it IEEE Trans.\ Nucl.\ Sci.} \ \issue({\bf #1},#2,#3)}
\def\IJMP(#1,#2,#3){{\it Int.\ J.\ Mod.\ Phys.} \ \issue({\bf #1},#2,#3)}
\def\JHEP(#1,#2,#3){{\it J.\ High Energy Physics} \ \issue({\bf #1},#2,#3)}
\def\JP(#1,#2,#3){{\it J.\ Phys.} \ \issue({\bf #1},#2,#3)}
\def\MPL(#1,#2,#3){{\it Mod.\ Phys.\ Lett.} \ \issue({\bf #1},#2,#3)}
\def\NP(#1,#2,#3){{\it Nucl.\ Phys.} \ \issue({\bf #1},#2,#3)}
\def\NIM(#1,#2,#3){{\it Nucl.\ Instrum.\ Meth.} \ \issue({\bf #1},#2,#3)}
\def\PL(#1,#2,#3){{\it Phys.\ Lett.} \ \issue({\bf #1},#2,#3)}
\def\PR(#1,#2,#3){{\it Phys.\ Rev.} \ \issue({\bf #1},#2,#3)}
\def\PRL(#1,#2,#3){{\it Phys.\ Rev.\ Lett.} \ \issue({\bf #1},#2,#3)}
\def\SJNP(#1,#2,#3){{\it Sov.\ J. Nucl.\ Phys.} \ \issue({\bf #1},#2,#3)}
\def\ZP(#1,#2,#3){{\it Zeit.\ Phys.} \ \issue({\bf #1},#2,#3)}
\def\be {\begin{equation}}
\def\ee {\end{equation}}
\def\bea {\begin{eqnarray}}
\def\eea {\end{eqnarray}}
\begin{document}
\begin{titlepage}
\begin{flushright}
CU-PHYSICS/06-2008 \hspace*{1.3in} HIP-2008-11/TH \hfill TIFR-TH-08-12
\end{flushright}
\begin{center}
{\Large {\bf 
{Universal Extra Dimensions, Radiative Returns and \\ [1mm] 
the Inverse Problem at a Linear $e^+ e^-$ Collider}}} \\[5mm]
\bigskip
{\sf Biplob Bhattacherjee} $^a$, 
{\sf Anirban Kundu} $^a$, 
{\sf Santosh Kumar Rai} $^b$, and
{\sf Sreerup Raychaudhuri} $^{c,}$\footnote{\scriptsize {\sl 
On leave of absence from the} {\rm Department of Physics, Indian 
Institute of Technology, Kanpur 208016, India.}}

\bigskip\bigskip

$^a${\footnotesize\rm 
Department of Physics, University of Calcutta, \\
92, Acharya Prafulla Chandra Road, Kolkata 700 009, India. \\
E-mail: {\sf bbhattacherjee@gmail.com, akphy@caluniv.ac.in} }

\bigskip
$^b${\footnotesize\rm 
Department of Physics, University of Helsinki and Helsinki Institute of 
Physics, \\ P.O. Box 64, FIN-00 014, University of Helsinki, Finland. \\
E-mail: {\sf santosh.rai@helsinki.fi} }

\bigskip
$^c${\footnotesize\rm 
Department of Theoretical Physics, Tata Institute of Fundamental Research, \\ 
1, Homi Bhabha Road, Mumbai 400 005, India. \\
E-mail: {\sf sreerup@theory.tifr.res.in} }

\normalsize
\vskip 10pt

{\large\bf ABSTRACT}
\end{center}

\begin{quotation} \noindent 
In a five-dimensional model with one Universal Extra Dimension (UED), 
signals for a pair of $n = 1$ Kaluza-Klein excitations could be easily 
observed at a future $e^+e^-$ collider if the process in question is 
kinematically allowed. However, these signals would prove difficult to 
distinguish from those predicted in other models, such as those with an 
extended gauge symmetry or supersymmetry. A much better power of 
discrimination is provided by the fact that the same machine could also 
produce the $n=2$ gauge bosons $\gamma_2$ and $Z_2$ as resonances in 
fermion pair-production without any upgrade in the collision energy 
$\sqrt{s}$. Assuming a fixed $\sqrt{s}$ -- as is expected at upcoming 
$e^+e^-$ machines -- we investigate the role of beam radiation in helping 
to excite such resonances through radiative returns. We then show how 
these resonances could yield unambiguous signals for UED, if taken in 
conjunction with the production of $Z_1$ pairs, identified by their 
decays to leptons and missing transverse energy.
\vskip 10pt
PACS numbers: {\tt 11.10.Kk, 12.60.Cn, 13.66.Hk} \\
\end{quotation}
\begin{flushleft}\today\end{flushleft}
\vfill
\end{titlepage}
\newpage
\setcounter{page}{1}
\section{Introduction}
\label{sec1}

\noindent The last decade has seen a remarkable revival of interest in 
phenomenological theories having extra spacetime dimensions -- an idea 
which goes all the way back \cite{ORaf} to N\o rdstrom (1914), Kaluza 
(1919) and Klein (1926). Today we have a number of such models, which 
differ on many specific counts, such as the number of extra dimensions, 
the geometry of space-time, and the choice of which fields to keep 
confined within the four canonical Minkowski dimensions while others are 
permitted to go into the extra dimensions or `bulk'. A generic feature 
of all these extra dimensional models is the fact that many of these 
detailed features are decided {\em ad hoc} rather than derived from some 
underlying principle -- the hope being that a more fundamental theory, 
when revealed, would provide the necessary dynamics. In this work, we 
focus on the Universal Extra Dimension (UED) model proposed by 
Appelquist, Cheng, and Dobrescu \cite{acd}, which has one extra compact 
dimension and {\em all} the fields of the Standard Model (SM) are 
defined over the bulk. In such models, the compact extra dimension(s) 
cannot form a simple manifold like the circle, sphere or torus 
considered in traditional Kaluza-Klein theories \cite {App}, because 
that would not be able to support the chiral fermions known to be 
present in the SM and hence would not permit parity-violation. A more 
exotic topology is, therefore, required. In the simplest UED scenario 
--- to which we confine ourselves --- there is only one extra dimension, 
denoted by $y $, but this is compactified on a $\mathbb{S}_1/ 
\mathbb{Z}_2$ orbifold, i.e.  a circle of radius $R$ folded about one of 
its diameters. Mathematically, this means that we simultaneously impose 
two symmetries, viz. $y \to y + 2\pi R$ and $y \to -y$. This 
`orbifolding' is sufficient to provide the necessary distinction between 
chiral components of fermions.

\bigskip\noindent The particle phenomenology of UED models is based on 
the following major features \cite{acd,cms1}:
\begin{itemize}
\item When projected in four spacetime dimensions, every bulk field 
$\Psi_P(x^\mu,y)$ is associated with a tower of Kaluza-Klein (KK) 
excitations $P_0, P_1, P_2, \dots$ The mass of $P_n$ is given, at the 
tree-level, by
\be
M_n^2 = M_0^2+ \frac{n^2}{R^2} \ , 
\label{kktree}
\ee
where $n = 0, 1, 2, \dots$ and $M_0$ is the mass of the SM particle $P_0$.

\item Since all the fields can access the bulk, momentum is expected to 
be conserved along the fifth compact direction, and hence the KK number 
$n$ is conserved. To get chiral fermions, we need the $\mathbb{Z}_2$ 
symmetry $y\to -y$ which breaks the translational symmetry along $y$ and 
induces terms located at the orbifold fixed points $y=0$ and $y=\pi R$. 
Obviously, such terms will violate KK-number conservation, though we can 
expect such effects to be suppressed by the boundary-to-bulk ratio. On 
the other hand, since there is no physical difference between the two 
fixed points at $y=0$ and $y=\pi R$, we obtain a residual $\mathbb{Z}_2$ 
symmetry $y\to y+\pi R$. The outcome, so far as the effective 
interaction Lagrangian after compactification goes, is the conservation 
of KK-parity -- defined as $(-1)^n$ -- at every interaction vertex.

\item The terms located at the orbifold fixed points are logarithmically 
divergent, i.e.  they are proportional to $\log \Lambda^2$ where 
$\Lambda \gg R^{-1}$ is the scale up to which the theory is expected to 
be valid.

\item Thus, the minimal UED model has two free parameters: $R$ and 
$\Lambda$, in addition to the mass $M_H$ of the SM Higgs boson. It is 
convenient to carry out phenomenological studies of the UED model with 
reference to the plane defined by $R^{-1}$ (dimension of mass) and 
$\Lambda R$ (dimensionless) as Cartesian axes. This constitutes the 
`theory space' of the UED model.

\item While the tree-level masses of the KK excitations follow the 
formula given in Eq.~(\ref{kktree}), important changes are observed at 
the one-loop level, where quantum numbers like spin, flavour and colour 
begin to play a role. In fact, these corrections control the 
mass-splittings and mixing angles between states (and hence allowed 
transitions) for any given excitation level with $n \neq 0$. Since 
many of the radiative corrections at the one-loop level are 
logarithmically divergent, this introduces a critical dependence of the 
phenomenological effects on $\Lambda$ at every level except $n = 0$.

\item Since KK-parity is conserved, the lightest $n=1$ particle must be 
stable.  In the minimal UED model, this is almost always the $\gamma_1$ --
the $n=1$ analogue of the photon. It turns out that this is more-or-less 
identical with the $n = 1$ excitation of the hypercharge gauge boson 
$B$, since the `Weinberg angles' for all $n \neq 0$ levels turn out to 
be small. The $\gamma_1$ is an excellent cold dark matter candidate -- 
in fact, this is one of the most attractive features of the UED model. 
In a collider experiment, this Lightest KK Particle (LKP) is `invisible' 
because of its weak interaction with matter\footnote{Because of 
conservation of KK parity, the interaction of a $\gamma_1$ with ordinary 
matter can only be by exchange of an $n = 1$ (or higher) particle, and 
hence the cross-section is suppressed by the large masses of these 
excitations. This is somewhat similar to the way in which 
neutrino-matter interactions are suppressed by the large mass of the 
exchanged $Z$ boson.}, leading to a characteristic signature with large 
amounts of missing energy and momentum.

\item All $n=1$ particles will undergo decays which eventually cascade 
down to the LKP. However, as the mass splitting among the $n=1$ states 
is generally small (being induced by radiative corrections), these 
cascade decays will generally yield one (or more) {\it soft} lepton(s) 
or jet(s) associated with a large missing $p_T$ due to the LKP. 

\end{itemize}

\noindent The close resemblance of this phenomenology with that of 
supersymmetry (SUSY) -- in its usual $R$-parity conserving avatar -- is 
immediately obvious. All $n = 0$ excitations are identical with the $R_p 
= 1$ particles of the SM, while all $n = 1$ excitations are analogous to 
the $R_p = -1$ super-partners.  Conservation of KK-parity plays the same 
role as conservation of R-parity: the stable LKP behaves exactly like 
the stable LSP (lightest SUSY particle) in leading to missing energy and 
momentum signals. UED signals could, therefore, easily be mistaken for 
SUSY signals and vice versa \cite{cms2}. It is true that in the case of 
UEDs, the spin of the $n > 0$ excitations is the same as that of their 
$n=0$ counterparts, but this is more a matter of detail, where the 
phenomenological effects are concerned. It is, therefore, a valid -- 
perhaps crucial -- question to ask if a new physics signal with the 
expected features is due to a underlying UED or to a SUSY theory, and to 
look for ways to confirm this.

\bigskip\noindent One important difference between SUSY and UED models 
immediately springs to the eye. This is the fact that in UED models, 
instead of one set of heavy partners of the SM particles, there is a 
whole tower of such partners for {\it each} SM particle. Thus, if we 
could pair-produce not just the $n=1$ KK modes, but also the $n=2$ KK 
modes, and find their masses to be in the expected ratio (approximately 
1:2, modulo radiative corrections), that would constitute very strong 
circumstantial evidence \cite{cms2} for UED. Pair-production of $n=2$ 
resonances would be a simple matter, if we had at our disposal a machine 
with arbitrarily large energy, but in practice, we only have machines 
with a limited kinematic access. In fact, the current experimental lower 
bounds on $R^{-1}$ of around 300--400 GeV already make it problematic to 
pair-produce the $n=2$ states at existing and planned colliders, 
including the Large Hadron Collider (LHC). One can still try to 
differentiate $n=1$ UED signals from those of SUSY models by trying to 
identify the spin of the intermediate states from the angular 
distribution of the decay products \cite{webber}. This could, however, 
be a tricky business, requiring the reconstruction of the rest frame of 
the decaying particles in the presence of large amounts of missing 
momentum.

\bigskip\noindent There is, however, one saving grace, and this is the 
fact that the $n=2$ modes can couple singly to a pair of the $n=0$ modes 
(i.e. SM particles), since all of them have positive KK-parity. It is 
thus possible, for example, to have $\gamma_2$ and $Z_2$ resonances in a 
four-fermion process, where both the initial and final di-fermion states 
are purely ($n = 0$) SM particles. Since the masses of these KK modes 
satisfy $M_{\gamma_2} \approx 2R^{-1} \approx 2M_{\gamma_1}$ and 
$M_{Z_2} \approx 2R^{-1} \approx 2M_{Z_1}$, the resonance energy is more 
or less the same as that required to {\it pair}-produce $\gamma_1$'s or 
$Z_1$'s. Thus, if one can reach this energy threshold, one could, in 
principle, obtain much stronger evidence for UED by demanding the {\it 
simultaneous} presence of these resonances with the signals from pair 
production of the $n = 1$ KK states.  The obvious place to look for such 
$s$-channel resonances would be in the LHC data \cite{asesh1}, which are 
expected to start coming in soon. Unfortunately, however, the $\gamma_2$ 
and $Z_2$ resonances can be shown to decay almost exclusively into 
jets\footnote{We note that the leptonic branching ratios of the $n=2$ 
excitations of $\gamma$ and $Z$ are much smaller than the leptonic 
branching ratios of an ordinary ($n=0$) $Z$ boson. It may still be 
interesting to carry out a search for the resonant contributions to 
Drell-Yan dileptons at the LHC, though the smallness of the excess 
cross-section could make such delicate searches difficult (and 
inaccurate) in the messy environment of a hadron collider \cite{Us}.} 
(without any missing energy). Any signal with a pair of jets is sure to 
be completely lost against the enormous QCD background at the LHC. One 
must, then, turn to a high-energy $e^+e^-$ collider, such as the 
proposed ILC, to observe these resonances in dijet production. Here, the 
production cross-section would be suppressed by the smallness of the 
$\gamma_2~e^+e^-$ or $Z_2~e^+e^-$ coupling, but this disadvantage can be 
largely offset by enhancement due to resonant effects and the high 
luminosity expected at such a machine. Obviously, dijet final states 
have a much smaller background at an $e^+e^-$ machine, making them 
viable for new physics searches.

\begin{figure}[h]
\setcounter{figure}{0}
\centerline{ \epsfxsize= 6.5 cm\epsfysize=10.0cm \epsfbox{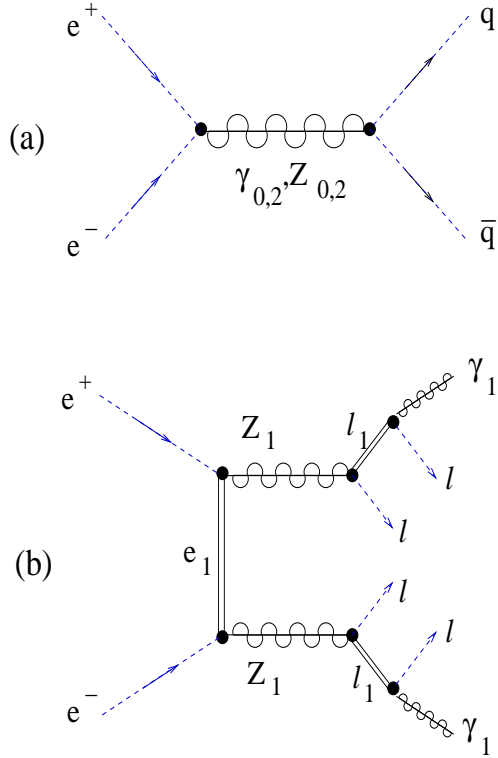} }
\caption{{\footnotesize\it Feynman diagrams showing the most important 
processes involving KK excitations at an $e^+e^-$ collider. The upper 
graph, marked {\em (a)}, illustrates $s$-channel exchange of $\gamma$, 
$Z$ and $\gamma_2$, $Z_2$, which add coherently. The lower graph, marked 
{\em (b)}, illustrates pair production of $Z_1$ excitations, which also 
has a crossed diagram (not shown in the figure). Leptonic decays of the 
$Z_1$ are also illustrated, with the symbol $l$ generically standing for 
any charged lepton or neutrino. Dashed (blue) lines indicate SM 
particles.}}
\label{fig:Feyn}
\end{figure}

\noindent In Figure~\ref{fig:Feyn}, we show some of the major Feynman 
diagrams contributing to UED signals at a high-energy $e^+e^-$ collider. 
The upper diagram, marked (a), shows dijet production with exchange of 
real or virtual KK excitations (bearing even KK number) of the photon or 
$Z$-boson, while the lower one, marked (b), shows pair-production of 
$Z_1$ excitations, leading to a final state with $4\ell + 
\not{\!\!E}_T$. An earlier study \cite{bk01} has shown that resonant 
production, as indicated in Figure~1($a$) is indeed a viable 
possibility, with strong resonances being obtained when the machine 
energy $\sqrt{s}$ coincides with the poles of the $\gamma_2$ and/or 
$Z_2$ propagators. However, the analysis in Ref.~\cite{bk01} makes the 
assumption that signals for UED would be discovered at the LHC, leading 
to an approximate knowledge of the masses of $\gamma_1$ and $Z_1$ --- 
and hence, of the compactification parameter $R^{-1}$. Knowing this 
parameter -- and therefore, the masses $M_{\gamma_2}$ and $M_{Z_2}$ -- 
it was assumed that the centre-of-mass energy $\sqrt{s}$ of the $e^+e^-$ 
collider would be {\it tuned} to these resonances, in the same way as 
LEP-1 had been tuned to the $Z_0$ resonance after its discovery at the 
UA1/UA2.

\bigskip\noindent In this context, one of the most often-quoted pieces 
of folklore in high energy physics is the statement that the next 
high-energy $e^+e^-$ collider --- presumably the planned International 
Linear Collider (ILC) --- will bear the same relation to the LHC as the 
LEP experiment bore to the UA1/UA2, i.e. if a resonant state is found at 
the LHC, precision measurements of that state would be made at the ILC. 
This is, in essence, the assumption made in Ref.~\cite{bk01}. However, 
there is a crucial difference between the planning of LEP and the ILC, 
viz. the fact that LEP-1 was tuned to the $Z$ resonance, which had 
already been discovered at the UA2, whereas the ILC is being planned 
even before the LHC has commenced its run.  According to this plan, the 
ILC will operate at {\it fixed} energies of $\sqrt{s} = 500$~GeV, with a 
later upgrade to $\sqrt{s} = 1$~TeV. It would, therefore, be entirely a 
matter of luck if the pole of a resonant state happens to lie at (or 
close to) these two values of the collision energy. A much more likely 
scenario, however, would have the resonance in question lying several 
decay widths away from the machine energy, in which case it will not 
show up as a signal of any significance. This is as true of the 
Kaluza-Klein resonances of the UED model as of other exotic particles 
like an extra $Z'$ or a massive graviton, and indeed of the SM $Z$ boson 
itself.

\bigskip\noindent All is not lost, however, for a way to observe these 
states will be provided by Nature herself. The discovery of resonant 
states far away from the machine energy $\sqrt{s}$ still remains 
possible, because of the well-known phenomenon of `radiative returns' at 
an $e^+e^-$ machine, which was first observed in the LEP-1.5 runs at 
$\sqrt{s} = 130$-136 GeV. There, the occasional emission of a hard 
collinear photon with energy $E_\gamma$ from the initial 
electron/positron state had led to a corresponding degradation of the 
effective centre-of-mass energy to $\sqrt{\hat{s}} = \sqrt{s} - 
E_\gamma$. During the LEP-1.5 run, there were a small number of such 
events where the energy was sufficiently reduced to match the $Z^0$-pole 
around $\sqrt{\hat{s}} = M_Z \simeq 91$~GeV, and the large resonant 
cross-section around the pole ensured that this extra contribution to 
the total $e^+e^-$ cross-section became quite substantial.

\bigskip\noindent In the LEP-1.5 runs, the radiation from the beam 
constituents, i.e.  electron/positron occurred because of the strong 
electromagnetic fields inside a bunch of electrons as it moved through 
the beam pipe. This effect, called {\em initial state radiation}, or 
ISR, is always present and can be predicted fairly accurately using QED 
alone. At the ILC, this ISR effect will certainly be present, but it 
will also be accompanied by another effect, viz. {\em beamstrahlung}, 
which is the emission of photons from the initial state 
electron/positron just before a collision under the influence of the 
electromagnetic fields of the other colliding bunch. This second QED 
effect was very weak at the LEP because the number density of charged 
particles in a bunch was rather low compared to the projected number 
density at the ILC. A low bunch density was made possible at the LEP 
because the design luminosity was easily attainable by multiple 
collisions (some 2000 per bunch) in the storage ring geometry. At a 
single-pass linear collider, however, the higher luminosity requirement 
demands tightly-packed bunches generating strong electromagnetic fields. 
To cut a long story short, therefore, radiative returns at the ILC will 
be a combined effect due to both ISR and beamstrahlung, and will create 
a substantial spread in the energy of the colliding beams\footnote{This 
unavoidable energy spread is among the reasons why the ILC design 
envisages a pre-determined machine energy rather than one tuned to a 
possible resonance.}. Turning this fact to our advantage, we can claim 
that even in the case of a fixed-energy ILC (or any other high-energy 
$e^+e^-$ machine), such radiative returns could well be the discovery 
mechanism of such resonances \cite{GoRaRaC}. The rarity of a radiative 
return (the probability is suppressed by the QED coupling $\alpha$) 
would be adequately offset by the large resonant cross-section at the 
pole of the propagator.

\bigskip\noindent Coming back to the UED model, we have already noted 
that the energy $E_{Z_1Z_1}$ required to pair-produce the $Z_1$ is 
around $E_{Z_1Z_1} \approx 2R^{-1}$, which is the same as the energy 
required for resonance production of $\gamma_2$ or $Z_2$, both of which 
have masses close to $2R^{-1}$. Thus, if $R^{-1}$ is small enough for 
these states to be accessible to an $e^+ e^-$ machine (radiative returns 
and all), it would be quite feasible to carry out a simultaneous study 
of the $4\ell + \not{\!\!E}_T$ signal from a $Z_1Z_1$ pair as well as 
the dijet signals from resonant $\gamma_2$ and $Z_2$ states. If we find 
a significant effect in {\it both} these channels, it can be argued 
quite convincingly that the circumstantial evidence is strong enough to 
pin down a unique signal for UED -- one which would discriminate it not 
only from the SM but also from SUSY, massive gravitons, extra gauge 
bosons, etc. in other rival models going beyond the SM.

\bigskip\noindent At this point, one may pause to reflect on the $Z_1$ 
pair production process and note that the dominant decay $Z_1 \to \ell 
\bar{\ell} \gamma_1$ includes both charged leptons and neutrinos under 
the generic symbol $\ell$. It is possible, therefore, for a $Z_1$ to 
decay into a $\nu \bar{\nu} \gamma_1$ final state, which would be 
completely invisible. There are, therefore, three distinct possibilities:
\begin{enumerate}
\item Both $Z_1$-s decay to charged lepton pairs and missing energy -- 
the signal has four charged lepton tracks with unbalanced (transverse) 
momentum;
\item One $Z_1$ decays to a pair of charged leptons and missing energy, 
the other decays invisibly -- the signal has two charged lepton tracks 
with unbalanced (transverse) momentum; and
\item Both $Z_1$-s decay invisibly. 
\end{enumerate} 
The last possibility may be immediately discounted, since there would be 
nothing to trigger on. We shall see, presently, that the leptons coming 
from the $Z_1$ decay are relatively soft, with $p_T < 100$~GeV in general. 
This means that for the second option, viz. a dilepton with missing 
energy, there will be a significant background from soft processes 
tending to produce dileptons, such as two-photon processes, vector boson 
fusion, etc. There will even be some background from decays of soft 
hadrons which would normally be dismissed as noise in the hadron 
calorimeter. It is best, therefore, to focus on the first option, viz. 
$e^+e^- \to \ell^+\ell^+\ell^-\ell^-+ \not{\!\!\!E}_T$, which would have 
relatively little soft background, and can be clearly distinguished from 
other hard processes by the low $p_T$ of the detected leptons.

\bigskip\noindent The thrust of this work is, therefore, twofold. The 
first part is to show that radiative returns caused by ISR and 
beamstrahlung effects can excite the $\gamma_2$ and $Z_2$ resonances in 
$e^+e^-$ collisions and provide observable effects, even when their 
masses are far away from the machine energy. The second part is to show 
how one can combine these with the signals for $Z_1$ pair-production at 
the same machine to identify a potential new physics discovery as 
specifically due to UED. As both dijets and $4\ell + \not{\!\!E}_T$ will 
certainly be among the final states which would be studied anyway when 
the ILC or any other high energy $e^+e^-$ collider becomes operational, 
the present study suggests an economical and accurate test for UED at 
such a machine.

\bigskip\noindent This article is organised in the following manner. In 
Section 2, we describe some details of the UED model which are relevant 
to the phenomenological studies which follow. This is followed in 
Section 3 by a brief introduction to the technique used to estimate the 
level of ISR and beamstrahlung. Our results on resonance production with 
radiative returns are presented and analysed in Section 4, while Section 
5 describes how this can be used in conjunction with $Z_1$ pair 
production to identify UED signals uniquely. Section 6 contains a brief 
summary and a few further comments of a general nature.

\vskip 10pt
\section{The UED Model}

\noindent It has already been mentioned that in UED models, all the SM 
fields live in a $\mathbb{R}_4 \times (\mathbb{S}_1/ \mathbb{Z}_2)$ 
spacetime, of which $\mathbb{R}_4$ is the canonical Minkowski space, and 
the remaining spatial dimension corresponds to a circle folded about a 
diameter. Naturally, all the particle momenta in the UED model will have 
five components, of which the component along the compact fifth spatial 
dimension has to be discrete, increasing in steps of size $R^{-1}$. 
Compactification to four dimensions then yields a Kaluza-Klein (KK) 
tower of states for every field, with the discrete fifth component of 
momentum appearing as an extra contribution to the mass $M_0$ of the 
zero mode, as shown in Eq.~(\ref{kktree}). Apart from the zero mode 
(identified with the SM particle), the other KK excitations are expected 
to be heavy if the parameter $R^{-1}$ is chosen large enough.

\bigskip\noindent The value of $R^{-1}$, i.e. the size of the extra 
compact dimension, is of paramount importance in UED phenomenology. 
Obviously, it cannot be too low, else there would already be 
experimental evidence for KK excitations. As a matter of fact, current 
experimental data already seem to prefer $R^{-1}$ to be in the range of 
a few hundreds of GeV. On the other hand, if $R^{-1}$ is too high, the 
KK excitations would be beyond the kinematic reach of upcoming collider 
experiments, a scenario which, though not impossible, would be very 
disappointing. Fortunately, however, there is a cosmological argument 
which militates strongly in favour of a value of $R^{-1}$ which would 
keep the lowest KK excitations within the range accessible to 
terrestrial experiments. We have already explained in the last section 
that KK-parity $(-1)^n$ has to be conserved and that this makes the LKP 
stable and weakly interacting, i.e. an excellent candidate for the cold 
dark matter (CDM) component of the Universe. Assuming, then, that the 
CDM is entirely composed of LKP's, we can take $\Omega_{\rm CDM} = 
0.110\pm 0.006~h^{-2}$ for the density of cold dark matter and obtain 
\cite{hooper} an upper bound\footnote{This can go down as far as 600 GeV 
in the eventuality that all three generations of SU(2) singlet $n=1$ 
leptons are nearly mass-degenerate (within 1\%) with the LKP $\gamma_1$.  
However, as this is only true for extreme choices of $\Lambda R$, we 
prefer the more conservative limit of 900 GeV.} on $R^{-1}$ of about 900 
GeV. This is well within the kinematic reach of planned and upcoming 
colliders -- of which the LHC is the prime example. However, it is only 
fair to say that this bound depends on two assumptions, viz. that there 
exists a substantial relic density of LKP-s and that there is no other 
component of cold dark matter. These assumptions are reasonable, but 
not, of course, absolutely essential. In this article, we accept them as 
binding and hence we focus on the ILC, which can explore up to about 
$R^{-1} = 450$~GeV and the CLIC, the proposed multi-TeV $e^+e^-$ machine 
at CERN, Geneva, which can explore the entire range up to $R^{-1} = 
900$~GeV.

\bigskip\noindent Complementing the cosmological upper bound, a number 
of low- and intermediate-energy processes constrain the parameter 
$R^{-1}$ at the lower end \cite{lowener,haisch,gogoladze}. For example, 
the non-observation of KK modes at colliding-beam machines like the LEP, 
Tevatron and HERA immediately tells us that the $n \neq 0$ excitations 
are beyond the kinematic reach of these machines, i.e. the lower bound 
on $R^{-1}$ is pushed up to at least 150~GeV. However, a much more 
stringent bound comes \cite{haisch} from the radiative decay $B\to 
X_s\gamma$. Taking both experimental error and theoretical uncertainties 
at the 5$\sigma$ level and making allowances for uncertain elements in 
QCD corrections to the new physics contribution, the current data on 
$B\to X_s\gamma$ roughly translates to a firm lower bound $R^{-1} \gtap 
~300$~GeV. Electroweak precision observables are also quite sensitive to 
$R^{-1}$, but the actual bound obtained is dependent on the mass $M_H$ 
of the Higgs boson. If $M_H$ is low ($\sim 115$~GeV), then the precision 
data severely restrict $R^{-1}$ to $R^{-1} ~\gtap ~600$~GeV at 99\% 
confidence level, but this bound will fall to about 400 GeV if the Higgs 
boson turns out to be heavy ($M_H \sim 350$~GeV) and even as low as 
300~GeV if $M_H$ is as large as 500~GeV \cite{gogoladze}. Taking the 
widest range into account, therefore, our study will consider $R^{-1}$ 
varying in the range $300~{\rm GeV} ~\ltap ~R^{-1} ~\ltap ~900~{\rm 
GeV}$.

\bigskip\noindent The formula for the mass of a KK excitation exhibited 
in Eq.~(\ref{kktree}) gets modified once we take into account the 
radiative corrections in a quantum field theory. Of course, like all 
such theories in higher dimensions, the UED model is non-renormalisable 
and must be treated in the spirit of an effective theory, valid up to a 
cutoff scale $\Lambda \gg R^{-1}$ --- at which point one must begin to 
consider an explicitly five-dimensional theory. Radiative corrections to 
the masses of the KK particles have been computed in Refs. \cite{cms1}, 
\cite{georgi} and \cite{pk}, where it has been shown that for any KK 
level ($n>0$), the almost-mass-degenerate spectrum resulting from 
Eq.~(\ref{kktree}) splits up due to such correction terms.  There are, 
in fact, two classes of corrections, viz. \begin{itemize}

\item The {\it bulk corrections}: These correspond to loop diagrams that 
are sensitive to compactification. The bulk corrections are small -- 
identically zero for fermions -- and hardly play an important role in 
determining the mass spectrum.

\item The {\it boundary corrections}: These terms are related with the 
interactions that are present only at the orbifold fixed points. They 
are much larger than the bulk corrections and are divergent, growing as 
$\log\Lambda$. It is these boundary corrections which play a major role 
in determining the exact spectrum (and hence possible decay modes) of 
the KK excitations for $n>0$.

\end{itemize} 

\vskip -3pt

\noindent Once the mass spectrum and the interactions are determined, 
one can proceed to make a phenomenological study of the UED model. 
Studies of the low-energy phenomenology of this model may be found in 
Refs. \cite{acd}, \cite{lowener}, \cite{haisch} and \cite{gogoladze}, 
while high-energy collider signatures are discussed in Refs. 
\cite{bk01}, \cite{lhc-ued}, \cite{susy-ued} and \cite{bk023}. Much of 
this work may be summed-up and absorbed in the choice of range $300~{\rm 
GeV} ~\ltap ~R^{-1} ~\ltap ~900~{\rm GeV}$ for the size of the compact 
dimension. To get concrete predictions we choose two benchmark values 
$R^{-1} = 400$~GeV and $R^{-1} = 800$~GeV, close to the upper and lower 
ends respectively, of this range, noting that both of these choices 
would predict KK excitations which could be accessible at planned 
$e^+e^-$ machines. The value of the cutoff scale $\Lambda$ is a somewhat 
tricky question, as it corresponds to the breakdown of the 
compactification limit -- which is, after all, a gradual process. Once 
again, we choose two benchmark values $\Lambda = 20R^{-1}$ and 
$50R^{-1}$ for our analysis, which correspond to an assumption that the 
low-energy effective theory will remain valid till a scale of some tens 
of TeV. As with most cutoff scales, the numerical value chosen is only a 
ballpark value, but this does not really matter as the dependence of the 
mass spectrum and couplings on $\Lambda$ is logarithmic, and hence not 
very sensitive to the precise number chosen. A visual presentation of 
the parameter space of this UED model may be found in 
Figure~\ref{fig:Correlation}.

\begin{table}[h]
\centering
\begin{tabular}{lcrrrr}
\hline
Benchmark Point: & & (I) & (II) & (III) & (IV) \\ \hline
Size Parameter & $R^{-1}$          &  400  & 400   & 800    & 800    \\
Cutoff Scale   & $\Lambda R$       &   20  &  50   &  20    &  50    \\
   \hline\hline
 $n = 1$ excitations 
         & ${\gamma_1}$      & 401.2 & 401.3 &  800.0 &  799.9 \\
         & ${E_1}$           & 404.4 & 405.7 &  808.7 &  811.4 \\
         & ${L_1}$           & 412.0 & 415.6 &  823.9 &  831.3 \\
         & ${W_1^\pm}$       & 430.8 & 437.6 &  850.4 &  864.2 \\
         & ${Z_1}$           & 431.4 & 438.1 &  850.5 &  864.3 \\ \hline
 $n = 2$ excitations 
         & ${\gamma_2}$      & 800.7 & 800.6 & 1599.7 & 1599.5 \\
         & ${L_2}$           & 818.4 & 825.7 & 1636.8 & 1651.4 \\
         & ${Z_2}$           & 840.2 & 854.1 & 1674.4 & 1702.4 \\
\hline\hline   
\end{tabular}	
\caption{\footnotesize\it Partial mass spectrum of the UED model at the 
four chosen benchmark points, showing only the lowest-lying states. All 
numbers are in GeV, except for $\Lambda R$, which is dimensionless.}
\label{tab:masses}
\end{table}

\noindent
In Table~\ref{tab:masses}, we list the masses of the KK excitations ($n 
= 1, 2$) relevant for the current study, for the four chosen benchmark 
points:
\bea
{\rm (I)}  && \! R^{-1} = 400~{\rm GeV}, \qquad 
\Lambda = 20R^{-1} = ~8~{\rm TeV} \ ; \nonumber \\
{\rm (II)} && \! R^{-1} = 400~{\rm GeV}, \qquad 
\Lambda = 50R^{-1} = 20~{\rm TeV} \ ;\nonumber \\
{\rm (III)} && \! R^{-1} = 800~{\rm GeV}, \qquad 
\Lambda = 20R^{-1} = 16~{\rm TeV} \ ; \nonumber \\
{\rm (IV)} && \! R^{-1} = 800~{\rm GeV}, \qquad 
\Lambda = 50R^{-1} = 40~{\rm TeV} \ .\nonumber
\eea
In every case, the $\gamma_1$ (partnering the massless photon $\gamma$) 
is the LKP, while the next-to-lightest $n=1$ excitation is the SU(2) 
singlet lepton $E_1$, followed by the SU(2) doublet leptons $L_1 = 
\left(\nu_1, \ell_1 \right)^T$.  For all practical purposes, the three 
generations of $E_1$ are degenerate, and a similar statement can be made 
about the $L_1$ as well. These states, together with the $W_1^\pm$ and 
the $Z_1$, all lie in the ballpark (within about 15\%) of $R^{-1}$ -- 
the splitting mostly coming from radiative corrections. In the $2R^{-1}$ 
regime, the relevant particles are $\gamma_2$, $Z_2$, and the SU(2) 
doublet lepton $L_2 = \left(\nu_2, \ell_2 \right)^T$, whose ${\mathbb 
Z}_2$-even components are left-chiral. The singlet $E_2$, whose 
${\mathbb Z}_2$-even component is right-chiral, though having a mass 
close to $2R^{-1}$, hardly couples to the $Z_2$ because the latter is 
almost completely dominated by the third component of the $W_2$ triplet. 
Other excitations, e.g. of quarks, are generically heavier than the 
corresponding ones shown in the table, because they have larger 
radiative corrections (for quarks, this is mainly due to large colour 
factors). Since this class of KK excitations is not relevant for our 
analysis, we do not exhibit the corresponding masses in 
Table~\ref{tab:masses}.

\bigskip\noindent At this point it is worth recalling the rule-of-thumb 
that if there is enough energy to pair-produce $n=1$ excitations, then 
there is also enough energy to excite a single $n=2$ resonance. Single 
production of $n=1$ states is, as we have seen, forbidden by the 
conservation of KK parity. Now, at the ILC-1, running at $\sqrt{s}=500$ 
GeV, no $n = 1$ KK excitation can be pair-produced, in view of the lower 
limit $R^{-1} ~\gtap ~300$~GeV, neither can the $n=2$ resonances be 
singly produced. Thus, the first phase of the ILC would be uninteresting 
for UED searches. At the ILC-2 --- the energy upgrade running at 
$\sqrt{s}=1$ TeV --- pair production of $n=1$ states as well as single 
production of $n=2$ resonances can take place if $R^{-1} < 500$ GeV. As 
this certainly holds for our first pair of benchmark points (I) and 
(II), we focus largely on the ILC-2 for our analysis. To study the 
heavier states that are predicted for 500~GeV~$<R^{-1}<900$ GeV --- 
where our second pair of benchmark points (III) and (IV) lie --- we 
shall need the CLIC, running at $\sqrt{s} =3$~TeV and 5 TeV, with a 
planned \cite{CLIC} luminosity of $10^{35}$~cm$^{-2}$~s$^{-1}$. This 
machine could, in fact, explore the complete $n=1$ spectrum as well as 
the $n=2$ resonances all the way up to the cosmological upper limit of 
$R^{-1}$ mentioned above.

\bigskip\noindent Since KK-parity is conserved, the possible decay modes 
for the low-lying KK states listed in Table~\ref{tab:masses} are 
severely restricted. A list of the important decay modes is given below. 
The decay channels important for this work are marked with a tic 
($\sqrt{}$) sign.
\vskip -15pt
$$
\begin{array}{cccl|cccl}
&\gamma_1    & : & {\rm stable~(invisible)}                           &
\sqrt{} &\gamma_2    & : & \gamma_2 \to f + \bar{f}                    \\
& E_1        & : & E_1 \to \ell + \gamma_1                            &  
& E_2        & : & E_2 \to \ell + \gamma_2                          \\
\sqrt{} & L_1        & : & \nu_1 \to \nu + \gamma_1, 
                  ~\ell_1 \to \ell + \gamma_1                         &  
& L_2        & : & \nu_2 \to \nu_1 + \gamma_1,
                  ~\ell_2 \to \ell_1 + \gamma_1                       \\
& W_1        & : & W_1 \to \nu + \bar{\ell}_1, ~\ell + \bar{\nu}_1     &&&    \\
\sqrt{} & Z_1        & : & Z_1 \to \nu + \bar{\nu}_1, ~\ell + \bar{\ell}_1     & 
\sqrt{} & Z_2        & : & Z_2 \to f + \bar{f}, ~\ell_1 + \bar{\ell}_1, 
                  ~\nu_1 + \bar{\nu_1}                                 \\
\end{array}
$$

\noindent Here, $f$ stands for any generic $n=0$ fermion (leptons as 
well as quarks) while $\ell$ and $\nu$ stand, respectively, for charged 
and neutral components of the SU(2) doublet leptons $L$. The SU(2) 
singlet (charged) leptons are denoted $E$. Generation indices have been 
suppressed and SM particles are written without the $n = 0$ subscript. 
It is important to note that the above list, limited as it is, contains 
both KK number-conserving as well as KK-number-violating processes. A 
generic feature of the UED model is that the branching ratios for the 
latter, though nonzero, are suppressed by the boundary-to-bulk ratio. 
The branching ratios for the KK number-conserving decays are, in turn, 
suppressed by the limited phase space available, because states with the 
same KK number have rather small mass-splittings. Eventually, therefore, 
both types of decay turn out to be of comparable importance. It is 
important to note that, unlike the $Z$-boson in the SM, the $Z_1$ decays 
overwhelmingly through the leptonic channels, i.e. either $Z_1 \to 
\ell^+\ell^- \gamma_1$, through a resonant $\ell_1^\pm$ (which decays to 
$\ell^\pm \gamma_1$ with unit branching ratio), or, $Z_1 \to 
\nu\bar{\nu} \gamma_1$, through a resonant $\nu_1$ (which decays to $\nu 
\gamma_1$, again with unit branching ratio). Since $n=1$ excitations of 
quarks are generically heavier than the $z_1$, its hadronic decays occur 
through three-body processes, and have partial widths that are at least 
three orders of magnitude smaller than their leptonic counterparts. 
Obviously, out of the 6 lepton flavours ($e, \mu, \tau$ and $\nu_e, 
\nu_\mu, \nu_\tau$) available in $Z_1$ decays, dilepton plus missing 
$E_T$ signals can be obtained only by counting the $e$, $\mu$, and 
$\tau$ flavours. This means that only 50\% of the $Z_1$-s produced at an 
$e^+e^-$ machine will lead to observable signals.

\begin{figure}[ht]
\centerline{ \epsfxsize= 5.7 cm\epsfysize=7.3cm \epsfbox{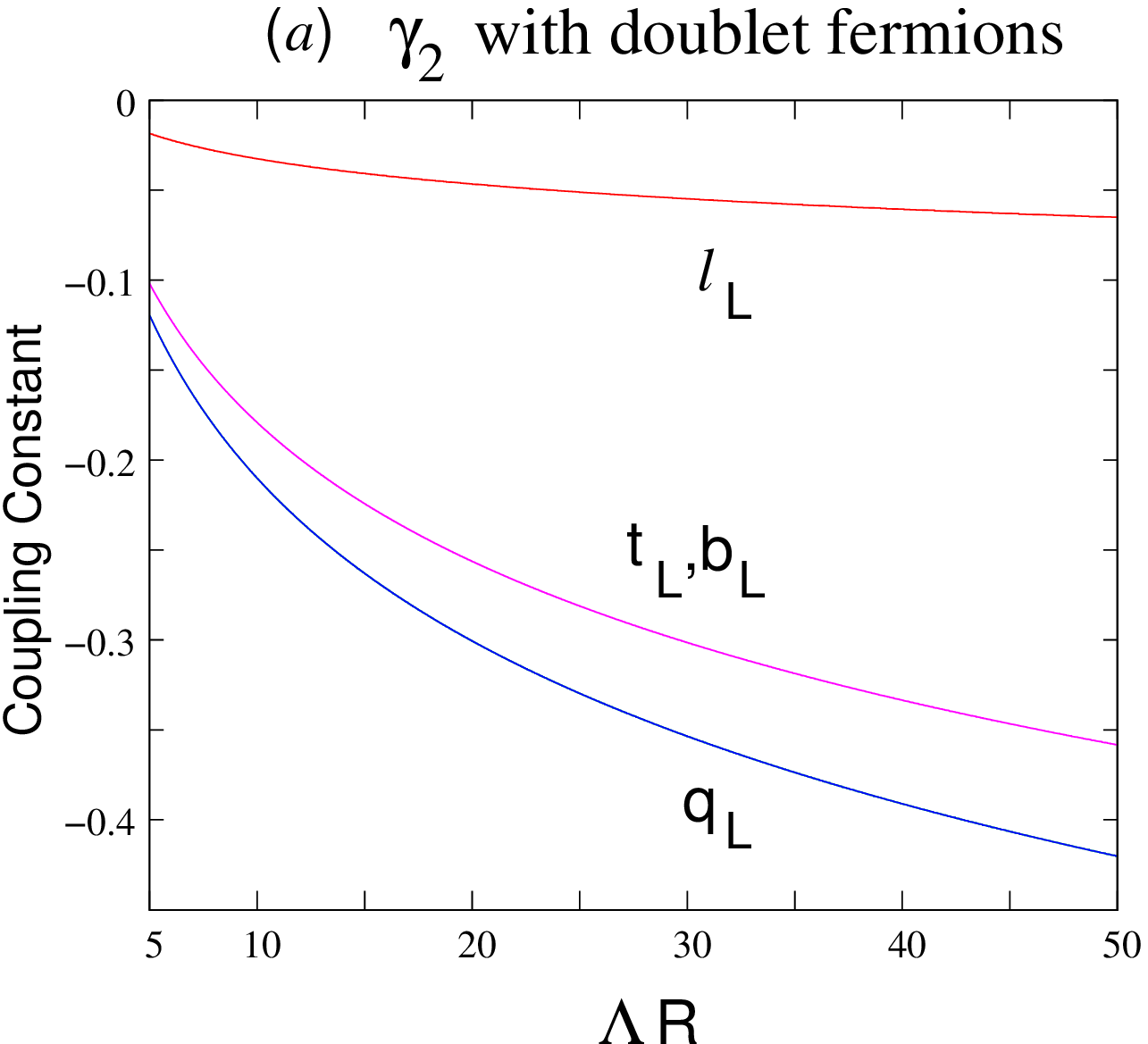} 
\hspace*{-0.2cm} 
\epsfxsize= 5.2 cm\epsfysize=7.3cm \epsfbox{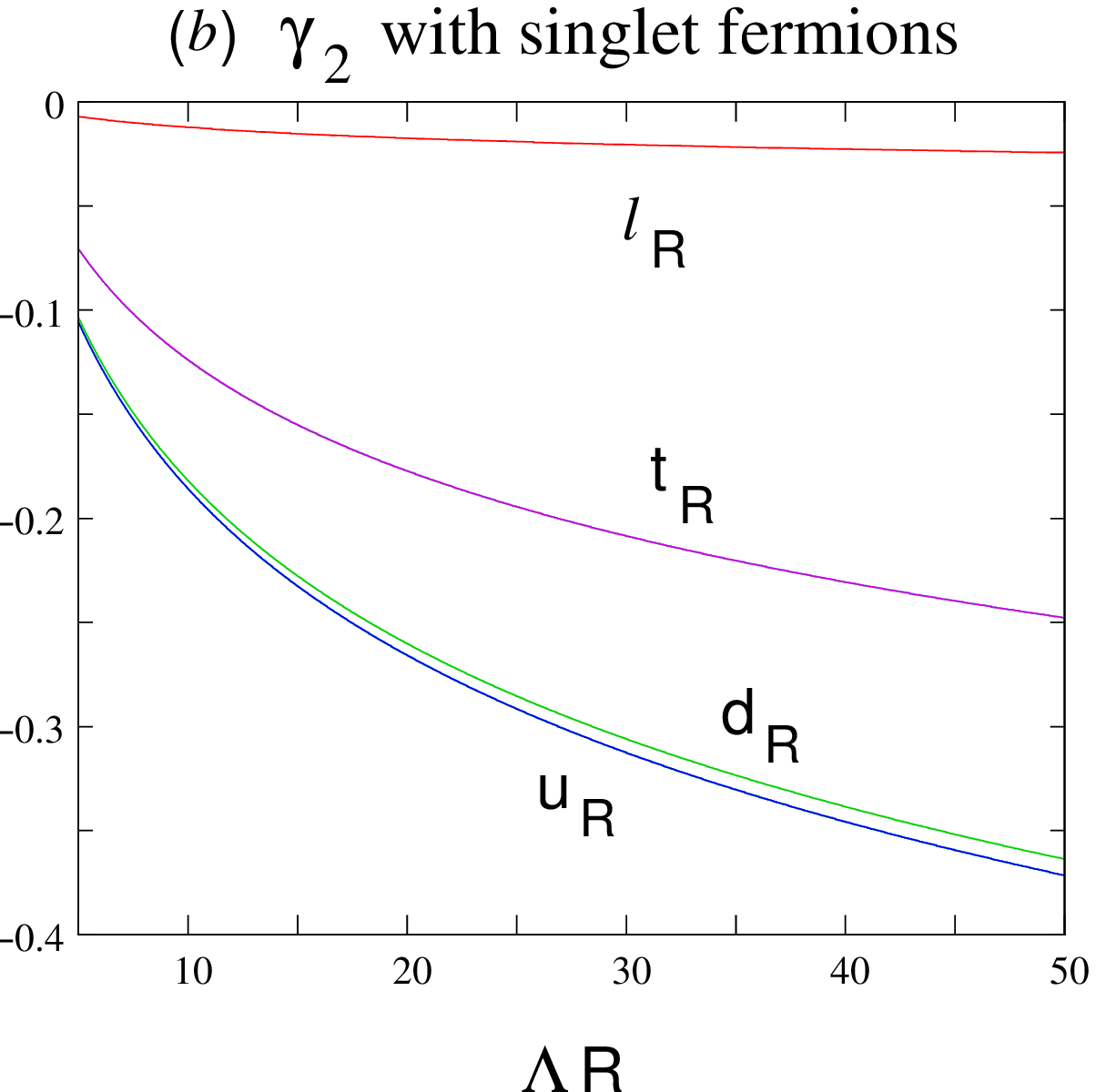} 
\hspace*{-0.2cm}
\epsfxsize= 5.3 cm\epsfysize=7.3cm \epsfbox{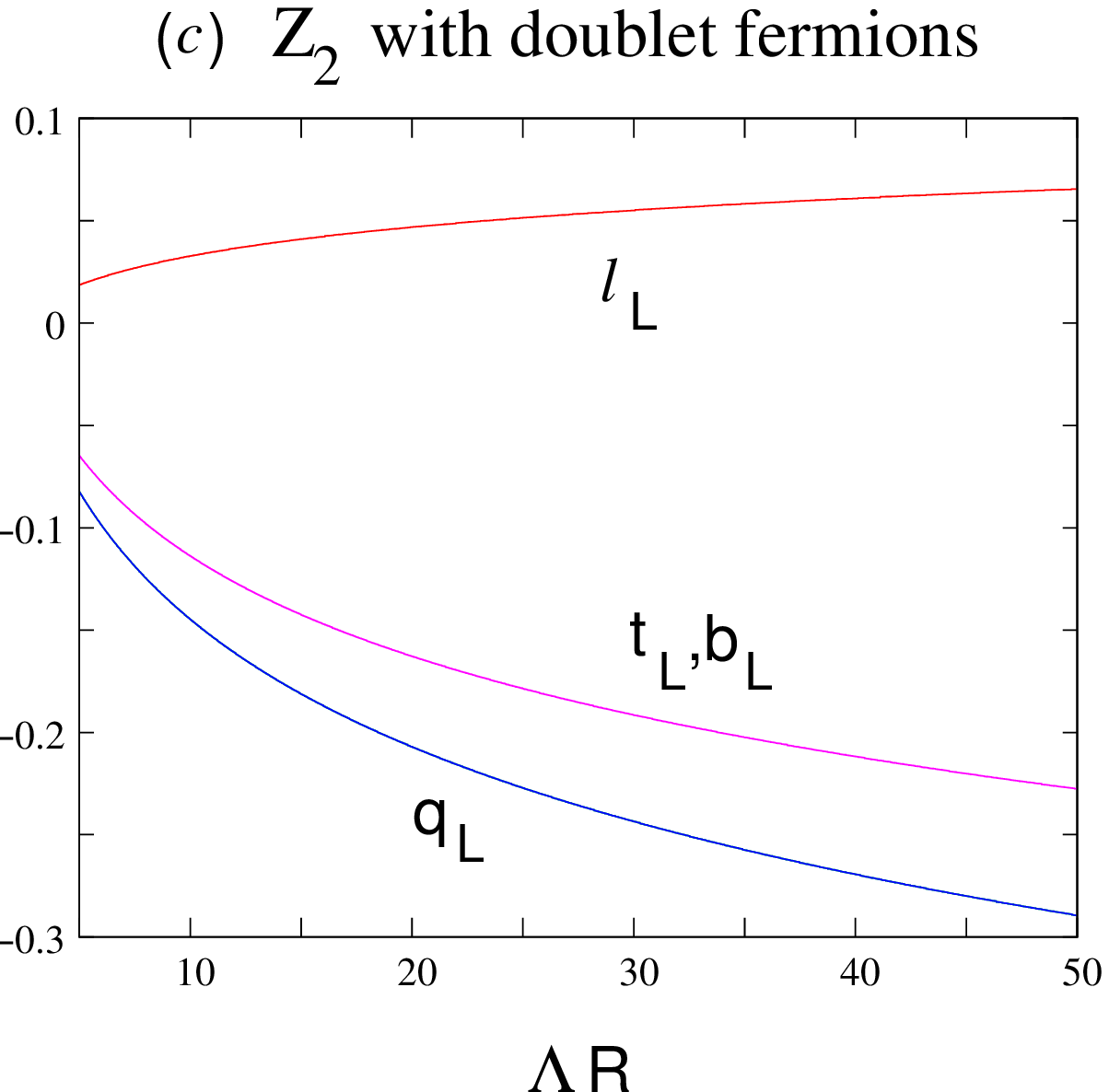} }
\vspace*{-0.2cm}
\caption{{\footnotesize\it Illustrating the variation in coupling of 
fermion pairs to the $\gamma_2$ and $Z_2$ excitations. Each box carries 
a header explaining the type of coupling, and each curve carries a 
legend explaining which fermions are involved. In every box, $l_{L,R}$ 
stands for any charged lepton $e$, $\mu$ or $\tau$, and $t_{L,R}$ stands 
for the top quark while $b_L$ indicates the bottom quark. In the boxes 
marked ($a$) and ($c$), $q_L$ stands for any light quark $u$, $d$, $s$ 
or $c$. In the box marked ($b$), $u_R$ includes $c_R$, while $d_R$ 
includes both $s_R$ and $b_R$. Note that the $Z_2 l_L \bar{l}_L$ 
coupling is positive, while all others are negative.}}
\label{fig:Coupling}
\end{figure}

\noindent To study the production and decay modes of the $\gamma_2$ and 
$Z_2$ resonances, we need to know how they couple to the SM ($n = 0$) 
fermions.  As the relevant formulae may be obtained from 
Ref.~\cite{cms1} or Ref.~\cite{bk01}, we have merely illustrated the 
results here. In Figure \ref{fig:Coupling}, we show the variation of the 
couplings of $n=0$ fermion-antifermion pairs with the $n=2$ electroweak 
gauge bosons as $\Lambda R$ varies from 5 to 50. The $\Lambda R$ 
dependence originates from the fact that these couplings conserve 
KK-parity, but violate KK-number and hence are very sensitive to 
radiative corrections.  A cursory examination of 
Figure~\ref{fig:Coupling} is enough to establish that the magnitudes of 
these couplings increase more or less as $\log\Lambda$. This is, of 
course, characteristic of an effective theory and is not incompatible 
with the Froissart bound. We also note the fact that at the $n=2$ level, 
the `Weinberg angle' quantifying the mixing between $W_2^{3\mu}$ and 
$B_2^\mu$ states is so small that it is quite reasonable to take the 
$Z_2^\mu$ as a pure $W_2^{3\mu}$, coupling only with doublet fermions, 
at least for all practical purposes. In fact, the tiny $Z_2f_R\bar{f}_R$ 
couplings hardly play any part in our discussions and hence we do not 
exhibit them with the others in Figure \ref{fig:Coupling}.

\bigskip\noindent It is interesting that the production cross-sections 
of the $n=2$ resonances are quite viable even though the coupling 
strength to an $e^+e^-$ pair, as exhibited in Figure~\ref{fig:Coupling}, 
is quite small ($\sim 10^{-2}$). This happens, of course, because of the 
resonance effect, where the coupling constant cancels out of the final 
cross-section in the narrow-width approximation. The resonant 
cross-section remains viable even when convoluted with the electron 
luminosity function $f_{e/e}(x)$. Naturally, this grows stronger as the 
machine energy and the resonant mass approach each other as $x \to 1$. 
On the other hand, in the same limit, the cross-section for producing a 
pair of $n=1$ excitations gets suppressed since the phase space gets 
squeezed towards zero volume. The phase space suppression factor, 
roughly $\left( 1 - 4/R^2s \right)^{3/2}$ is, in fact, enough to 
neutralise the larger coupling of $n=1$ states to SM fermions, so that, 
eventually, the pair-production cross-section becomes somewhat smaller 
than the dijet cross-section. In addition, when we trigger only on final 
states with charged leptons, there is an additional suppression factor 
of $\frac{1}{2} \times \frac{1}{2}$, which reduces the cross-section to 
a quarter. Kinematic cuts further suppress this cross-section, as 
explained in Section~4. Despite all these suppression factors, however, 
the high luminosity expected at the ILC or CLIC serves to keep the 
$4\ell + \not{\!\!E}_T$ signal viable. Since this is one of the cleanest 
signals we can have, with a very small SM background, it is well worth 
including in our study, as we shall see presently.
  
\bigskip\noindent In this section, then, we have enlisted the details of 
the UED model that are relevant for our numerical analysis of the 
problem and for prediction of the relevant signals. In the next section, 
we describe how to include radiation effects and obtain an effective 
spectrum which can be convoluted with the formulae in Ref.~\cite{bk01} 
to get realistic predictions, using the choices of free parameters 
delineated in the present section.

\vskip 10pt
\section{ISR and Beamstrahlung}

To study the effect of ISR and beamstrahlung in electron (positron) 
beams, we make use of the so-called structure function formalism 
\cite{Drees}. Assuming that an initial electron (positron) of energy 
$E_b$ emits a photon of energy $E_\gamma$, leaving an electron 
(positron) of energy $E_e$, we define the energy fraction $x = E_e/E_b$ 
so that $E_\gamma = (1 - x) E_b$. We now define a normalised probability 
distribution $f_{e/e}(x)$ for the colliding electron (positron) to have 
energy $E_e = xE_b$. This quantity is analogous to the parton density 
function $f_{q/p}(x)$ or $f_{g/p}(x)$ at a hadron collider. Hence, if we 
consider the process $e^+ (p_1)~e^-(p_2) \to q~\bar q~(\gamma)$, the 
cross-section will be
\begin{equation}
\sigma\left[ e^+(p_1) e^-(p_2) \to q \bar q (\gamma) \right]
= \int dx_1~dx_2~ f_{e/e}(x_1) ~f_{e/e}(x_2) ~~\hat{\sigma}\left[e^+(x_1p_1) 
e^-(x_2p_2) \to q \bar q\right]
\end{equation}
where $\hat{\sigma}$ denotes the cross-section calculated with the 
(degraded) initial momenta $x_1p_1$ and $x_2p_2$. The effective 
centre-of-mass energy is $\sqrt{\hat{s}} \approx \sqrt{x_1 x_2 s}$, 
where $s = (p_1 + p_2)^2$ and $\hat{s} = (x_1p_1 + x_2p_2)^2$ in the 
high energy limit when the electron (positron) mass can be neglected.

\bigskip\noindent The electron `luminosity' function $f_{e/e}(x)$ is a 
combination of the probabilities for both ISR and beamstrahlung effects 
and it may be calculated \cite{Drees} by convoluting the corresponding 
(normalised) spectral densities as follows:
\begin{equation}
f_{e/e}(x) = \int_x^1 \frac{dy}{y} ~f_{e/e}^{\rm ISR}(y) ~f_{e/e}^{\rm beam}
\left(\frac{x}{y}\right)
\label{eqn:convolute}
\end{equation} 
For the ISR spectral density $f_{e/e}^{\rm ISR}(y)$, we use a one-loop 
corrected Weiz\"acker-Williams approximation to write
\begin{equation}
f_{e/e}^{\rm ISR}(y) = \frac{\omega}{16} 
\left[ (8 + 3\omega)(1 - y)^{\omega/2 - 1} - 4(1 + y) \right]
\label{eqn:ISR}
\end{equation}
where 
\begin{equation}
\omega = \frac{2\alpha}{\pi} \left( \log\frac{s}{m_e^2} - 1 \right)
\end{equation}
with $\alpha$ and $m_e$ denoting the fine-structure constant and the 
electron mass respectively, both evaluated at the scale $E_b$. Because 
of its weak logarithmic dependence on $s$, $\omega$ stays confined in a 
small range around $0.14\pm0.02$ for the energies under consideration. 
As a result the exponent in Eq.~(\ref{eqn:ISR}) is always negative, 
indicating that there will be a steep rise in $f_{e/e}^{\rm ISR}(y)$ as 
$y \to 1$. The normalisation condition $\int_0^1 dy ~f_{e/e}^{\rm 
ISR}(y) = 1$ ensures that there is no singularity, but the large values 
near $y = 1$ indicate that ISR effects are not very effective in 
spreading out the energy $E_e$ much below $E_b$.

\bigskip\noindent The formula for the spectral density due to 
beamstrahlung is much more complicated, as it depends critically on the 
(dimensionless) beamstrahlung parameter $\Upsilon$. This is given, for 
$e^-$($e^+$) beams with a Gaussian energy profile, by \cite{Chen}
\begin{equation}
\Upsilon = \frac{5r_e^2}{6\alpha m_e} \ 
\frac{E_b N_e}{\sigma_z (\sigma_x + \sigma_y)}
\end{equation}
where $r_e \approx 2.8\times 10^{-15}$~m is the classical electron 
radius, $\sigma_{x,y,z}$ are the dimensions of a bunch (assuming an 
ellipsoidal shape) and $N_e$ is the number of electrons in a bunch. The 
spectral density for multiple photon emission can be approximated 
\cite{Chen}, for values of $\Upsilon$ less than about 10, by the formula
\begin{equation}
f_{e/e}^{\rm beam} (\xi) = \frac{1}{N_\gamma} 
\left[\delta(1 - \xi) \left( 1 - e^{-N_\gamma} \right) + 
\left(1 - \xi + \xi\sqrt{1 + \Upsilon^{2/3}} \right)
\frac{e^{-\eta(\xi)}}{1 - \xi} \sum_{r=0}^\infty 
\frac{\eta(\xi)^{r/3}\Gamma_{r+1}(N_\gamma)}{r!~\Gamma\left(\frac{r}{3}\right)}
\right]
\end{equation}
where 
$$
N_\gamma = \frac{5\alpha^2m_e\sigma_z}{2r_eE_b}
\frac{\Upsilon}{\sqrt{1 + \Upsilon^{2/3}}}
$$
is the number of photons emitted per electron and $\eta(\xi) = 2(1 - 
\xi)/2\Upsilon \xi$. The symbol $\Gamma(x)$ stands for the usual Euler 
gamma function while $\Gamma_s(x)$ denotes the incomplete gamma function 
defined by
$$
\Gamma_s(x) = \int_0^s dt ~t^{1+x} e^{-t} \ .
$$

\noindent This formula is too complicated to reveal much to an 
inspection except, clearly, the important role played by the $\Upsilon$ 
parameter, but, when plotted as a function of the argument $\xi$, it 
does have a shape rather similar to that of the ISR spectrum -- except 
that the peak around $\xi = 1$ is not quite so sharp. However, what 
matters for our analysis is neither the ISR spectrum $f_{e/e}^{\rm 
ISR}(y)$ alone, nor the beamstrahlung function $f_{e/e}^{\rm beam} 
(\xi)$, but the convolution of the two shown in 
Equation~(\ref{eqn:convolute}), which will describe the actual energy 
spread.  Some of the important parameters required to generate these 
spectra are given below, in Table~2.

\begin{table}[h]
\centering
\begin{tabular}{ccccc}
\hline
Machine & $\sqrt{s}$~(TeV) & $\sigma_z$~(mm) & $\omega$ & $\Upsilon$ \\
\hline\hline 
ILC-1   & 0.5              & 0.30            & 0.1235  & 0.048      \\
ILC-2   & 1.0              & 0.30            & 0.1300  & 0.110      \\
CLIC    & 3.0              & 0.03            & 0.1402  & 8.100      \\
\hline\hline   
\end{tabular}	
\caption{\footnotesize\it Some of the beam parameters for the ILC and 
CLIC which are crucial to the evaluation of radiation spectra involving 
ISR and beamstrahlung.}
\label{tab:Beam}
\end{table}

\begin{figure}[h]
\centerline{ \epsfxsize= 8.0 cm\epsfysize=7.0cm \epsfbox{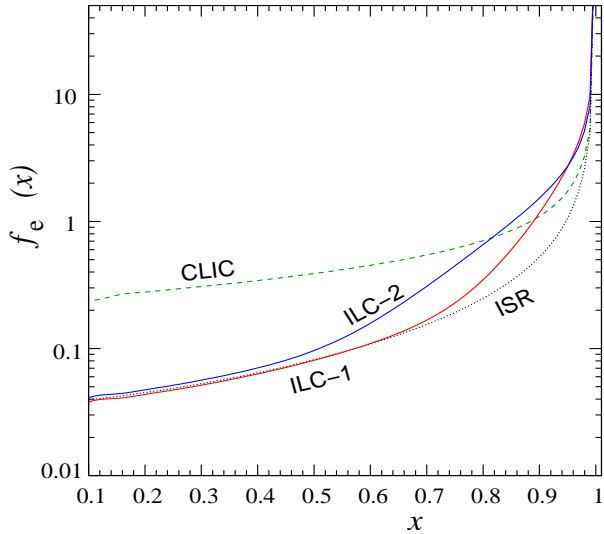} }
\vskip -10pt
\caption{{\footnotesize\it Effective electron (positron) flux at a high 
energy collider taking into account both ISR and beamstrahlung effects. 
Note the steep rise as $x \to 1$ and also the fact that the spreading 
effect is much larger at the higher energy of CLIC.}}
\label{fig:Flux}
\end{figure}

\noindent In Figure~\ref{fig:Flux} we have shown the behaviour of this 
`luminosity' function as the momentum fraction $x$ of the colliding 
electron (positron) varies from 0 to 1. The solid (red, blue) lines show 
the behaviour at the ILC-1 and ILC-2, running at $\sqrt{s} = 500$~GeV 
and $\sqrt{s} = 1$~TeV respectively, while the dashed (green) line shows 
the corresponding curve at the CLIC, running at 3~TeV. The dotted 
(black) line shows the ISR spectrum at the ILC-2 and is given 
essentially for purposes of comparison.

\bigskip\noindent The above graph allows one to make some quick 
estimates of the size of the resonance effects that are induced by 
radiative returns. For example, if the resonance occurs when the energy 
of the colliding particles is around 60\% of the machine energy, then we 
may expect $x \simeq 0.75$, where the luminosity at the ILC is around 
1\% of the value expected in the vicinity of $x \to 1$. This means that 
for both $e^+$ and $e^-$ combined, we get an effective flux of the order 
of $10^{-4}$. For reasonably sharp resonances, this small flux, 
compounded with the typical resonant cross-section, which is at least 
$10^4$ times the off-resonance cross-section (as the decay widths of 
$\gamma_2$ and $Z_2$ are in the ballpark of 1~GeV), would predict at 
least as large a contribution as that which we would predict in the 
absence of the radiative effects. For relatively lighter resonances, the 
effect is more pronounced at the CLIC, as Figure~\ref{fig:Flux} readily 
shows. Even apart from the contribution to the total cross-section, 
which is already considerable, we will get a more dramatic effect if we 
look at the invariant mass distribution of the particles produced 
through the resonant state. The next section discusses this issue in 
more detail.

\vskip 10pt
\section{Bump Hunting}

\noindent In order to make a numerical analysis and illustrate the 
effect of radiative returns, we have incorporated the electron 
luminosity function of Eq.~(\ref{eqn:convolute}) in a simple Monte Carlo 
event generator where we calculate

($a$) the dijet production process $e^+ + e^- \to 
\gamma^*/Z/\gamma_2/Z_2 \to q + \bar{q}$, and

($b$) the pair-production process $e^+ + e^- \to Z_1 + Z_1 \to \gamma_1 
\ell^+\ell^- + \gamma_1 \ell^+\ell^-$,

\noindent which have been illustrated in Figure~\ref{fig:Feyn}. In case 
($a$), the final state will be a pair of jets, and we focus on the 
invariant mass distribution of these jets, where distinct peaks 
corresponding to resonant $\gamma_2$ and $Z_2$ states should be seen 
over the SM background arising from $\gamma^\ast$ and $Z^\ast$ 
exchanges. In case ($b$), the final state will be 
$\ell^+\ell^+\ell^-\ell^-$ and a large missing $E_T$, where the four 
leptons can be either electrons (positrons), muons (anti-muons) or taus 
(anti-taus).  As a check, we have also incorporated the UED fields and 
couplings as a separate kernel in the software CalcHEP \cite{Pukhov} and 
generated the same signals, using the ISR and beamstrahlung generators 
built into this software. In every case, the numbers obtained from 
CalcHEP turn out to be in very close agreement with those obtained from 
our simulation codes.

\begin{figure}[h]
\centerline{\epsfxsize=13.0 cm\epsfysize=10.8cm \epsfbox{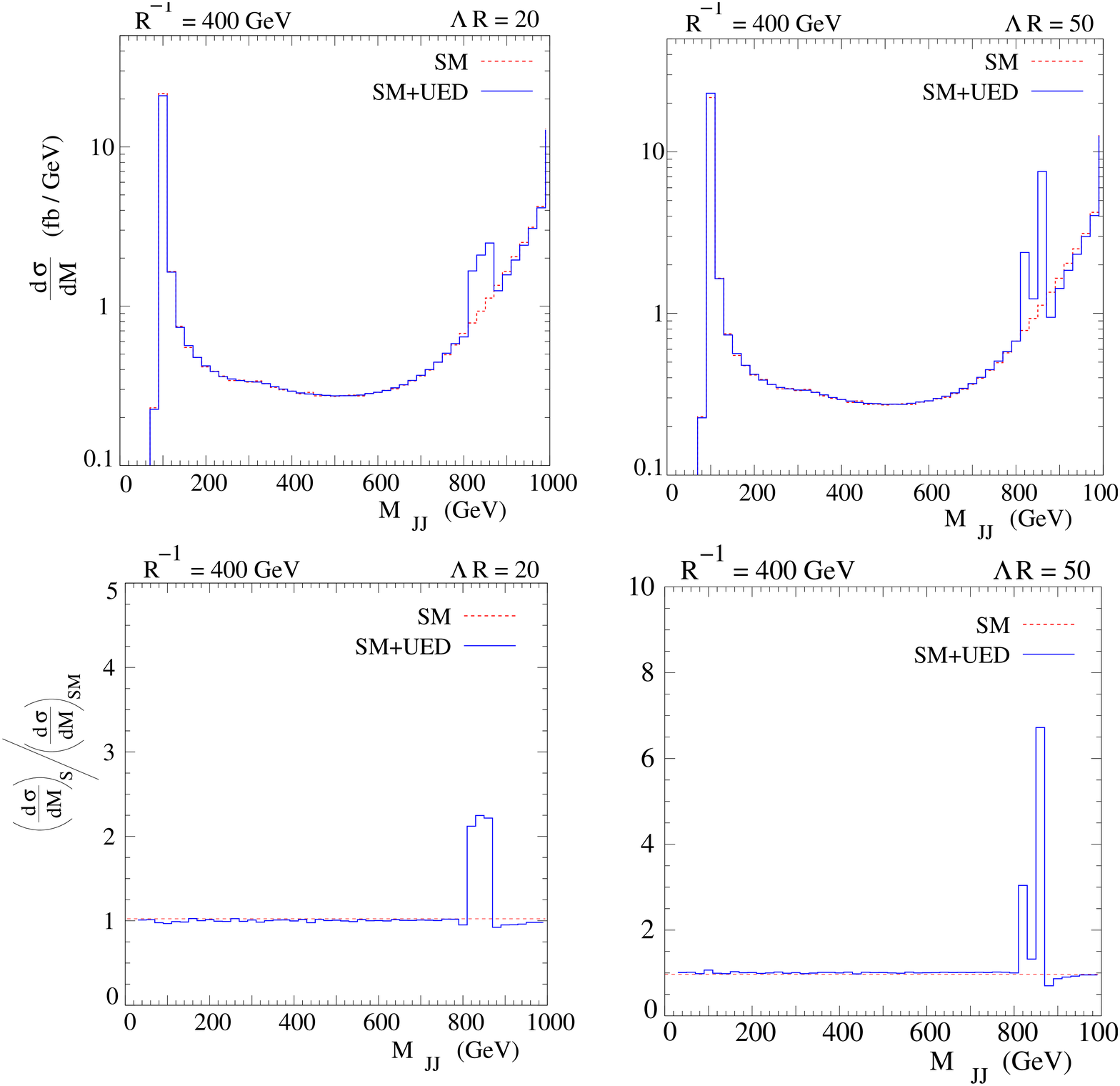} }
\caption{{\footnotesize\it Illustrating $\gamma_2$ and $Z_2$ resonances 
at the ILC-2, running at $\sqrt{s} = 1$~TeV. In the upper boxes, the 
dashed (red) histograms correspond to the SM prediction, while the solid 
(blue) histograms represent the effect of UED signals for $R^{-1} = 
400$~GeV and $\Lambda R = 20$ (left) and 50 (right). In the lower boxes, 
the same cross-sections are normalised by the SM prediction, effectively 
throwing the new physics effect into relief.}}
\label{fig:ILC}
\end{figure}

\bigskip\noindent As discussed before, we do not consider the ILC-1 for 
this analysis, because mass scales as large as $2 \times 400$~GeV are 
not kinematically accessible at a 500~GeV machine. Our results for the 
distribution in dijet invariant mass at the ILC-2 ($\sqrt{s} = 1$~TeV) 
are illustrated in Figure~\ref{fig:ILC}. For these plots, we have chosen 
the benchmark points (I) and (II) with $R^{-1} = 400$~GeV and $\Lambda R 
= 20$ and 50 respectively, where the $\gamma_2$ and $Z_2$ resonances are 
light enough to be produced on-shell. We plot the invariant mass 
distribution as a histogram with bin-size 20 GeV. This bin width is 
dictated by a simple-minded estimate of the detector-smearing effects on 
the measurement of the jet momentum. As these measurements will be 
calorimetric, the error is dominated by the error in the energy 
measurement, which is estimated \cite{ILC_TDR} as $\delta E_J \approx 
\pm 0.3\sqrt{E_J}$. Considering two jets of the same energy with errors 
adding in quadrature\footnote{This is consistent with the construction 
of the invariant mass in terms of the momenta (energies).}, the error in 
dijet invariant mass arising from energy measurements alone can be 
estimated as $\delta M_{JJ} \approx \sqrt{2}\times\delta E_J \approx \pm 
0.42\sqrt{E_J}$. For the ILC-2, running at $\sqrt{s} = 1$~TeV, $E_J 
\simeq 500$~GeV, which means that $\delta M_{JJ} \approx \pm 9.4$~GeV. 
If we add on the smaller errors due to thrust axis and angle 
measurements, we can expect an error of around $\delta M_{JJ} \approx 
\pm 10$~GeV. Thus, it is reasonable to choose a bin width of 20~GeV.

\bigskip\noindent To identify the jets, we impose the following 
acceptance cuts: 
\begin{itemize}

\item The pseudo-rapidity $\eta_J$ of the jets should satisfy 
$\mid\eta_J\mid \leq 2.5$;

\item The jet transverse momentum $p_T^{(J)}$ should satisfy $p_T^{(J)} 
\geq 10$~GeV.

\end{itemize}

\noindent These represent minimal requirements for identification of 
jets in the detectors that would be built to operate at ILC energies and 
means that we are considering essentially the entire phase space 
available in practice. We also require, however, to ensure that the 
dijets triggered on are being produced strictly from beam-beam 
collisions, and for this, we impose a selection cut:
\begin{itemize} 
\item The missing transverse momentum $\not{\!p}_T$ should satisfy 
$\not{\!p}_T \leq 10$~GeV. 
\end{itemize} 
The choice of 10~GeV for these selection and acceptance cuts on the jets 
is consistent with the uncertainty in jet energy described above.

\bigskip\noindent Since this work is essentially of exploratory nature, 
our event generator works only at the parton level, i.e. we identify the 
jet thrust axis and the jet energy-momentum with the direction and 
magnitude of the four-momentum of the parent quark. No simulation of the 
fragmentation of these quarks is attempted. For an $e^+e^-$ collider, 
this is known to be a reasonably good approximation when averaged over a 
large number of events, though an individual event may occasionally have 
very different characteristics. The analysis may, therefore, be carried 
out fairly accurately without using more sophisticated jet simulation 
algorithms, provided there is enough luminosity to yield a large 
number of events. The advantage of using a fast parton-level code is 
that it readily allows exploration of the parameter space beyond the 
four benchmark points chosen earlier. This advantage makes itself felt 
in the last section, where we discuss the entire accessible parameter 
space rather than a few points.

\noindent Our results for the invariant mass distribution are exhibited 
in Figure~\ref{fig:ILC}. The upper and lower boxes on the left show the 
distribution in invariant mass of the dijet in the final state for the 
benchmark points (I) and (II), i.e. $\Lambda R = 20$ and 50, 
respectively. The solid (blue) histogram shows the effect of the UED 
signal added to the SM background, while the dashed (red) histogram -- 
just about visible below the peak(s) -- represents the SM alone. As 
expected, the overall shape of the SM histogram reproduces the 
luminosity curve\footnote{Strictly speaking, the convolution of two such 
luminosity curves -- one for the $e^-$ and one for the $e^+$} in 
Figure~\ref{fig:Flux}, with the $x \to 1$ peak corresponding to the 
machine energy of 1~TeV. Instead of falling off at low $\sqrt{s}$ 
values, i.e. low $x$, however, the curve rises to a sharp peak in the 
bin $M_{JJ} = 90 - 110$~GeV, which represents a `return' to the 
$Z$-resonance, similar to what was observed at the LEP-1.5. The 
resonance is nowhere near as strong as it was at LEP-1.5 because the 
electron flux, as shown in Figure~\ref{fig:Flux} is quite low (around 
4\%) at $x = M_Z/\sqrt{s} \simeq 0.09$. At the extreme left end, the 
cross-section may be seen to drop rather abruptly towards zero --- this 
is not a dynamical effect, but an artefact of the acceptance cuts on the 
jets.

\bigskip\noindent The (blue) signal histograms in Figure~\ref{fig:ILC}, 
i.e. when the effects of $n=2$ resonances are included, correspond 
closely to the SM for most of the range in $M_{JJ}$, except for the 
resonant peak(s) around 800 -- 850 GeV, which represent(s) on-shell 
production of the $\gamma_2$ and $Z_2$ modes. If $\Lambda R = 20$, the 
two peaks, which are separated by around 40-50 GeV at the centres (see 
Table~\ref{tab:masses}), cannot be resolved with a realistic binning in 
the invariant mass spectrum, but if $\Lambda R = 50$, we obtain two 
distinct bumps. These are highlighted in the lower boxes, which plot the 
ratio of the signal histogram to the background histogram. The ratio is 
unity for most of the invariant mass range, except at the resonant 
peaks, where the excess stands out decisively. Small dips in the signal 
below the SM background, just observable to the right of the peaks, 
correspond to interference effects.

\bigskip\noindent
The following points may now be noted in the context of these resonances.
\begin{enumerate}

\item The $\gamma_2$ and $Z_2$ resonances are narrow ($\Gamma \sim 
1$~GeV) and sharp\footnote{The actual decay widths for the $\gamma_2$ 
($Z_2$) in GeV are 0.23~(0.51), 0.45~(0.98), 0.46~(0.98) and 0.90~(1.91) 
for the benchmark points I, II, III and IV respectively.}, corresponding 
to at least a twofold increase in the cross-section in the relevant 
bin(s), even when multiplied by the small flux factor. Note that the 
upper left box in Figure~\ref{fig:ILC} shows a single smeared-out 
resonance because the illustrated bump actually contains {\it two} 
unresolved resonances. As a matter of fact, we have checked that a finer 
binning of 10~GeV, were it experimentally possible, would have been 
enough to resolve the $\gamma_2$ and $Z_2$ peaks clearly even for 
$\Lambda R = 20$. Unfortunately, as we have discussed above, this would 
not be achievable at the ILC for $\sqrt{s} = 1$~TeV.

\item Since the excess appears only in one or two bins -- and even then 
the peaks are not very high -- the total dijet cross section will show a 
rather modest deviation from the SM. In order to reduce the SM 
background, therefore, we may impose some further selection cuts. A 
glance at Figure~\ref{fig:ILC} shows that a cut of $M_{JJ} > 600$~GeV 
would effectively reduce the SM background, without in the least 
affecting the resonances appearing in the 800--900~GeV ballpark. Such a 
cut is called for anyway, since we know that $R^{-1} > 300$~GeV, i.e. no 
resonances are expected below $2 \times 300$~GeV. Similarly, a cut on 
$p_T^{\rm jet} > 100$~GeV would remove all vestiges of the Jacobian peak 
arising from $Z$-decay, without significantly affecting the Jacobian 
peaks from the decays of the much heavier $\gamma_2$ and $Z_2$, even 
when we take smearing effects (due to spread in beam energy) into 
account. These selection cuts are not needed to discover resonances, but 
will be needed for the studies of the next section, where we show how 
the underlying model can be identified unambiguously.
   
\item Though we have not considered $e^+e^- \to \ell^+\ell^-$ processes 
in this work because of the small branching ratios of the $\gamma_2$ and 
$Z_2$ to leptonic channels, it may still be worth considering these 
channels, since the lepton momenta can be measured much more accurately, 
allowing for a finer binning of $\ell^+\ell^-$ invariant mass, and hence 
a clear resolution between the two resonant peaks for all values of 
$\Lambda R$. If, indeed, a clear signal of the kind predicted in this 
work is observed, then the resolution of the single peak into two 
closely-separated peaks might serve to clinch the issue of whether the 
underlying model incorporates UED or not. However, as the cross-section 
is small for $\ell^+\ell^-$ final states, such a result will be possible 
only with a large amount of statistics, and it may take all the data 
accumulated in the full run of the ILC-2 before we can reach any such 
conclusion. For this reason, we do not make any analysis of the 
$\ell^+\ell^-$ final states in this work.

\end{enumerate}

\begin{figure}[ht]
\centerline{ \epsfxsize=13.5 cm\epsfysize=13.0cm \epsfbox{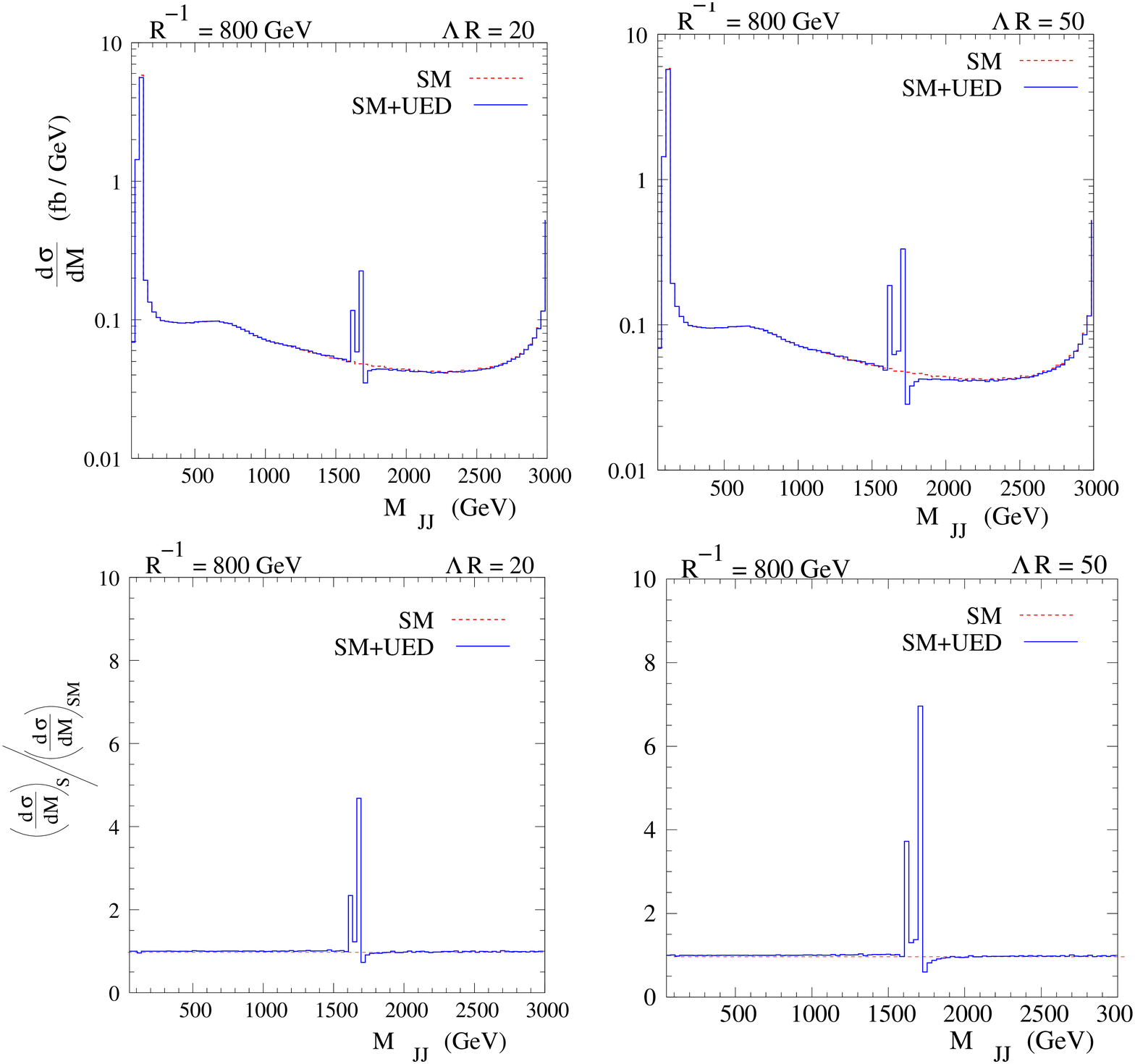} }
\caption{{\footnotesize\it Illustrating $\gamma_2$ and $Z_2$ resonances 
at the CLIC, running at $\sqrt{s} = 3$~TeV. All conventions are the same 
as in Figure~\ref{fig:ILC}, except that the parameter $R^{-1}$ is chosen 
to be 800~GeV and the bins are each 30~GeV wide.}}
\label{fig:CLIC}
\end{figure}

\noindent We have already remarked that even when the $\gamma_2$, $Z_2$ 
resonances in Figure~\ref{fig:ILC} add up to a single bump, the height 
of this is rather moderate, as resonances go. The situation is 
significantly improved if we go from the 1~TeV ILC-2 to the 3~TeV CLIC, 
where the large design value of $\Upsilon$ would be responsible for 
creating a much wider energy spread. This is illustrated in 
Figure~\ref{fig:CLIC}, where the notations and conventions of 
Figure~\ref{fig:ILC} are repeated. To avoid losing out too heavily on 
the low flux at low $x \approx 2R^{-1}/\sqrt{s}$, as well as to 
illustrate the superior kinematic reach of this higher energy machine, 
the value of $R^{-1}$ has been chosen as 800~GeV, rather than 400~GeV, 
for this figure, pushing the resonance energy to the neighbourhood of 
1.6-1.7~TeV. The bin-width has been increased to 30~GeV, which is compatible 
with jets of energy 1.5~TeV each and errors in energy measurement 
similar to those at the ILC.  A comparison of the the two figures will 
immediately show that at the CLIC, the resonant effect is clearer than 
at the ILC, the bump rising two to three times higher than the SM 
background in the most relevant bin. This greater height is due to a 
combination of three effects, viz. ($i$) the greater energy spread due 
to beamstrahlung at the CLIC (see Fig.~\ref{fig:Flux}), ($ii$) the 
larger $s$-channel suppression for the background at the higher energy 
of the CLIC, and ($iii$) the greater bin size. It is worth noting that 
at higher values of $R^{-1}$, such as have been chosen for 
Figure~\ref{fig:CLIC}, the separation between the $\gamma_2$ and $Z_2$ 
peaks is large enough to be clearly distinguishable, in spite of the 
increased bin size. The fact that the resonance(s) occur in the middle 
of the available range for $M_{JJ}$ also shows that the CLIC will have 
ample energy to scan the entire range of $R^{-1}$ which is permitted by 
the dark matter constraint.

\bigskip\noindent We reiterate the fact that a $Z_1$ pair and a single 
$Z_2$ (or $\gamma_2$) are kinematically accessible for the {\it same} 
machine energy, indicating that there will always be a {\it 
simultaneous} excess over the SM prediction in the cross-sections for 
$e^+e^- \to$~dijets, as well as in $e^+e^- \to 4\ell + \not{\!\!E}_T$. 
The presence of both these excess contributions would constitute a 
strong signal for new physics of the type considered in this work. 
Production and decay of $Z_1$ pairs at an $e^+e^-$ collider has been 
discussed in the literature \cite{Rizzo}, and hence we do not elaborate 
on the signal characteristics in this paper. As explained in a previous 
section, we concentrate on the dominant leptonic decay modes of the 
$Z_1$, viz.
$$
Z_1 \to \ell + \bar{\ell}_1 \to \ell + (\bar{\ell} + \gamma_1) \ ,
$$  
leading to a $4\ell + \not{\!\!E}_T$ signal at the ILC or CLIC. This can 
take place in two ways: $Z_1 \to \ell \bar{\ell}_1 \to \ell (\bar{\ell} 
\gamma_1)$ and its charge-conjugate process $Z_1 \to \bar{\ell} \ell_1 \to 
\bar{\ell} (\ell \gamma_1)$, both leading to the same final state 
$\ell\bar{\ell} + \not{\!\!E}_T$. The branching ratios are the same for 
all lepton flavours in the massless limit. Since there are six leptons 
($\ell = e,\mu,\tau,\nu_e,\nu_\mu,\nu_\tau$), the branching ratio for 
this process to each individual flavor must be $\frac{1}{6}$. These 
branching ratios require to be convoluted with the detection 
efficiencies to get a more realistic estimate. As stated before, we are 
interested only in the charged leptons $e$, $\mu$ and $\tau$. The 
efficiency factors may safely be taken \cite{ILC_TDR} as 95\% for the 
$e$ and $\mu$, and 85\% for the $\tau$. Armed with this information, we 
can analyse the different possibilities for a $4\ell + \not{\!\!E}_T$ 
final state as shown in Table~\ref{tab:branching}.

\begin{table}[h]
\centering
\begin{tabular}{crrrcl}
\hline
Channel & $Z_1 \to$~~~         & $Z_1 \to$~~~           & Final State~~~        
& B.R.                   &  Efficiency Factor \\ 
\hline\hline              
1 & $e^+e^-+ \not{\!\!E}_T$  & $e^+e^-+ \not{\!\!E}_T$  & $e^+e^+e^-e^-+ \not{\!\!E}_T$
& $\frac{1}{6} \times \frac{1}{6}$  &  $\times(0.95)^4$ \\
2 & $e^+e^-+ \not{\!\!E}_T$  & $\mu^+\mu^-+ \not{\!\!E}_T$  & $e^+e^-\mu^+\mu^-+ \not{\!\!E}_T$
& $\frac{1}{6} \times \frac{1}{6}$  &  $\times(0.95)^4$ \\
3 & $e^+e^-+ \not{\!\!E}_T$  & $\tau^+\tau^-+ \not{\!\!E}_T$  & $e^+e^-\tau^+\tau^-+ \not{\!\!E}_T$
& $\frac{1}{6} \times \frac{1}{6}$  &  $\times(0.95)^2\times(0.85)^2$ \\
\hline   
3 & $\mu^+\mu^-+ \not{\!\!E}_T$  & $e^+e^-+ \not{\!\!E}_T$  & $e^+e^-\mu^+\mu^-+ \not{\!\!E}_T$
& $\frac{1}{6} \times \frac{1}{6}$  &  $\times(0.95)^4$ \\
4 & $\mu^+\mu^-+ \not{\!\!E}_T$  & $\mu^+\mu^-+ \not{\!\!E}_T$  & $\mu^+\mu^+\mu^-\mu^-+ \not{\!\!E}_T$
& $\frac{1}{6} \times \frac{1}{6}$  &  $\times(0.95)^4$ \\
5 & $\mu^+\mu^-+ \not{\!\!E}_T$  & $\tau^+\tau^-+ \not{\!\!E}_T$  & $\mu^+\mu^-\tau^+\tau^-+ \not{\!\!E}_T$
& $\frac{1}{6} \times \frac{1}{6}$  &  $\times(0.95)^2\times(0.85)^2$ \\
\hline   
6 & $\tau^+\tau^-+ \not{\!\!E}_T$  & $e^+e^-+ \not{\!\!E}_T$  & $e^+e^-\tau^+\tau^-+ \not{\!\!E}_T$
& $\frac{1}{6} \times \frac{1}{6}$  &  $\times(0.95)^2\times(0.85)^2$ \\
7 & $\tau^+\tau^-+ \not{\!\!E}_T$  & $\mu^+\mu^-+ \not{\!\!E}_T$  & $\mu^+\mu^+\tau^-\tau^-+ \not{\!\!E}_T$
& $\frac{1}{6} \times \frac{1}{6}$  &  $\times(0.95)^2\times(0.85)^2$ \\
8 & $\tau^+\tau^-+ \not{\!\!E}_T$  & $\tau^+\tau^-+ \not{\!\!E}_T$  & $\tau^+\tau^+\tau^-\tau^-+ \not{\!\!E}_T$
& $\frac{1}{6} \times \frac{1}{6}$  &  $\times(0.85)^4$ \\\hline\hline
\end{tabular}	
\caption{\footnotesize\it Branching ratio (B.R.) and detector efficiency factors for 
different channels in $Z_1$ pair decay.}
\label{tab:branching}
\end{table}
\noindent Multiplying out the factors in every row and adding up all the 
rows leads to an overall suppression factor of about 0.17, instead of 
the 25\% we would have got if we had taken all the efficiencies to be 
unity.
 
\bigskip\noindent Finally, before we move on to a more general 
discussion, we need to note a couple of technicalities in the 
calculation of these excess contributions, viz.

\begin{itemize}

\item For the dijet cross-section, we should look for an identifiable 
resonance, such as the ones shown in Figures~\ref{fig:ILC} and 
\ref{fig:CLIC}. If the resonance effect is not apparent on inspection -- 
as it is in these figures -- some numerical criterion may be used 
instead, such as a $3\sigma$ deviation from the SM prediction in a pair 
of adjacent bins. It must be said, however, that within the minimal UED 
formalism, we will generally have very clear resonances, unless, for 
some reason, we have to take a very broad bin size. \item For the $4\ell 
+ \not{\!\!E}_T$ signal, we should expect the leptons to be relatively 
soft, since the mass gap between the parent $Z_1$ and the daughter 
$\gamma_1$ is not very large. A glance at Table~\ref{tab:masses} shows 
that at least for the four benchmark points, this splitting is never 
more than 35~GeV. The analogue of this may not be true for other kinds 
of new physics which also lead to a $4\ell + \not{\!\!E}_T$ final state. 
Hence, it is reasonable to impose a selection cut $p_T^\ell \leq 40$~GeV 
on the lepton transverse momentum at the ILC-2, where the parent $Z_1$-s 
are produced with very low $p_T$. At the CLIC, however, we will require 
to raise this cut, since now the $Z_1$-s will themselves carry 
considerable $p_T$. We find that a cut of $p_T^\ell \leq 100$~GeV is 
optimal at the CLIC, as it will hardly affect the signal in a UED model, 
but can affect the signal in other new physics models quite 
dramatically.

\item For the $4\ell + \not{\!\!E}_T$ signal, again, one must impose an 
acceptance cut of $p_T^\ell \geq 5$~GeV (10~GeV) at the ILC (CLIC), 
since the detectors will obviously not respond to very soft leptons. 
Unlike the upper cut, this will lead to a significant reduction in the 
signal cross-section, as the splitting between $n=1$ mass states can 
sometimes be rather low ($\sim 5 - 10$~GeV). However, as this cut is a 
requirement originating from the hardware constraints, we must perforce 
live with the corresponding loss in signal events. \end{itemize}

\noindent The numerical analysis in the following section has been done 
incorporating these selection criteria. We shall presently see their 
efficacy as a device to eliminate not merely the Standard Model 
backgrounds, but also the so-called `new physics backgrounds' at the 
ILC.

\vskip 10pt
\section{Tackling the Inverse Problem}

\noindent
We now address the culminating issue, which may be framed as a question: 

\begin{quotation}
\noindent {\it Would the observation of a high-mass resonance, or a pair 
of such resonances, at a high energy $e^+e^-$ collider tell us clearly 
that a universal extra dimension exists?}
\end{quotation}

\noindent This is a classic example of the so-called `inverse problem' 
at any high energy machine, i.e. the process of identifying a new 
physics effect as due to a specific model and a particular set of value 
of the model parameters. For the case in hand, the answer to the above 
question appears to be negative, since it can easily be argued that 
neither the discovery of resonant peak(s) in the $e^+e^- \to JJ$ 
invariant mass spectrum, nor the observation of an excess in $4\ell + 
\not{\!\!E}_T$ events, is by itself sufficiently distinctive to be 
touted as a `smoking gun' signal of UED. For example, a single bump in 
the dijet invariant mass spectrum could be interpreted as any one of the 
following:

\begin{enumerate}
\item A resonant $Z'$ boson, predicted in four-dimensional 
models with extra $U(1)$ symmetries.
\item A heavy sneutrino $\widetilde{\nu}_\mu$ or $\widetilde{\nu}_\tau$ 
in a SUSY model with $R$-parity-violating couplings of both $LL\bar E$- 
and $LQ\bar D$-type.
\item A massive graviton $G_1$, predicted in the Randall-Sundrum model 
with a warped extra dimension and two $D_3$-branes.
\end{enumerate}

\noindent If {\it two} resonances are clearly resolvable, the 
possibility of this signal being due to the massive Kaluza-Klein 
excitation $G_1$ of a graviton is more-or-less ruled out. However, one 
could very well have two nearly-degenerate $Z',Z''$ bosons in a model 
with two extra $U(1)$ symmetries. Similarly, one could have 
nearly-degenerate sneutrinos, $\widetilde{\nu}_\mu$ and 
$\widetilde{\nu}_\tau$ coupling to both $e^+e^-$ pairs as well as quarks 
with $R$-parity-violating couplings of similar strength. Of course, it 
may be possible to identify the spin of the exchanged particle 
\cite{GoRaRaC} by reconstructing the angular distribution of the jets in 
the centre-of-mass frame, though this would be hampered by the missing 
longitudinal momentum due to the radiated photon(s). If this difficulty 
can be overcome and the angular distribution indicates an exchanged 
particle of spin~1, the sneutrino and graviton options would be 
eliminated, but we would still have the $Z'$ option to reckon with.

\bigskip\noindent If we consider the $4\ell + \not{\!\!\!E}_T$ signal, 
standing alone, this can also have many new physics sources. The SM 
background arises primarily from final states with on-shell production 
of weak gauge bosons, such as $e^+ e^- \to W^+W^-Z \to (\ell^+\nu) 
(\ell^-\bar{\nu}) (\ell^+\ell^-)$, or from $e^+ e^- \to ZZZ \to 
(\ell^+\ell^-) (\ell^+\ell^-) (\nu \bar{\nu})$, where, in principle, 
each boson can decay into leptons of different flavour. There will also 
be a large number of sub-dominant diagrams contributing to the same 
six-lepton final states. As all of these can be accurately predicted in 
the SM, however, an excess in $4\ell + \not{\!\!E}_T$ events would be 
readily identified. Once such an excess is identified, however, it could 
be interpreted as due to any of the following alternatives, apart from a 
UED model.

\begin{enumerate}

\item A pair of heavy $Z'$ bosons, with an ordinary $Z$ boson radiated 
from any of the fermion legs, i.e. the process $e^+e^- \to Z'Z'Z$. We 
would require two of these neutral bosons to decay into $\ell^+\ell^-$ 
pairs, while the third decays invisibly to neutrinos. Generally, leptons 
coming from the decay of a $Z'$ boson would tend to be hard, and would 
fall foul of the $p_T^\ell \leq 40~(100)$~GeV cut imposed at the ILC-2 
(CLIC). If we further focus on the fact that the $\ell^+\ell^-$ 
invariant mass will peak at the mass of the parent boson -- unlike the 
case for a $Z_1$ decay in the UED model, where the leptons are part of a 
three-body decay -- it may be possible to distinguish this case from the 
UED case. Obviously, this construction will require a lot of statistics, 
which may be available only towards the end of the ILC run. A similar 
possibility is to produce a pair of $Z'$ bosons in association with a 
$\ell^+\ell^-$ pair, i.e. $e^+e^- \to Z'Z'\ell^+\ell^-$, which can arise 
from several diagram topologies, such as, e.g., $e^+e^- \to 
Z'Z'\gamma^*$ or $e^+e^- \to \ell^+\ell^-$ with two radiated $Z'$-s. One 
of the $Z'$ bosons would then have to decay to neutrinos, while the 
remaining one decays to another $\ell^+\ell^-$ pair (which will exhibit 
the invariant mass peak). Yet another possibility is $e^+e^- \to 
ZZ'\ell^+\ell^-$ (for which again there are several diagram topologies) 
which can be treated similarly.

\item Another alternative is to produce a pair of heavy $W'^\pm$ bosons, 
with an ordinary $Z$ boson radiated from any of the fermion legs, i.e. 
$e^+e^- \to W'^+W'^-Z$. Each $W'$ would decay to $\ell \nu_\ell$, while 
the $Z \to \ell^+\ell^-$ (all daughter pairs having, in principle, 
different lepton flavours). Such heavy $W'$s are predicted in many 
theories, such as left-right symmetric models, 3-3-1 gauge models, and 
so on. As before, the leptons will tend to be hard and lead to 
elimination of these events by the $p_T^\ell$ cut, and it may again be 
possible to distinguish this case from the UED one by vetoing the events 
when an $\ell^+\ell^-$ invariant mass peaks around $M_Z$ when enough 
statistics has been accumulated. An alternative process with $e^+e^- \to 
W'^+W'^-\ell^+\ell^-$ can also be envisaged, where, for example, the 
$\ell^+\ell^-$ pair arises from an off-shell photon radiation. 
Obviously, in this case, there may be no identifiable peak in the 
$\ell^+\ell^-$ invariant mass, as a result of which the signal would 
closely resemble the $4\ell + \not{\!\!E}_T$ signal from UED in all 
kinematic characteristics. A large part of this will be removed by the 
$p_T^\ell$ cut, but there may still be some irreducible part which 
cannot be distinguished from the UED signal.

\item A pair of heavy neutralinos $\widetilde{\chi}_i^0 
\widetilde{\chi}_j^0$ ($i,j>1$), each of which decays as 
$\widetilde{\chi}_i^0 \to \ell \widetilde{\ell} \to \ell (\ell 
\widetilde{\chi}_1^0)$, where $\widetilde{\ell}$ denotes a slepton. In a 
SUSY model where $R$-parity is conserved, the LSP $\widetilde{\chi}_1^0$ 
is invisible, so that we would eventually observe four leptons together 
with missing energy and momentum. If the mass gap between the 
$\widetilde{\chi}_i^0$ ($i>1$) and the $\widetilde{\chi}_1^0$ is not 
very large, the final state leptons can be expected to be reasonably 
soft -- mimicking the UED signal almost perfectly. This, in fact, is an 
example of the similarities because of which the UED model was 
originally nicknamed `bosonic SUSY'.

\item A pair of neutralinos $\widetilde{\chi}_i^0 \widetilde{\chi}_j^0$ 
($i,j=1,4$), each of which decays as $\widetilde{\chi}_i^0 \to \ell 
\widetilde{\ell}^* \to \ell (\ell \nu)$ or as $\widetilde{\chi}_i^0 \to 
\nu \widetilde{\nu}^* \to \nu (\ell \ell)$ in a SUSY model where 
$R$-parity is violated through $LL\bar E$ couplings. A similar 
possibility would be pair-production of charginos 
$\widetilde{\chi}_i^\pm \widetilde{\chi}_j^\mp$ ($i,j=1,2$), where one 
decays to $\widetilde{\chi}_i^\pm \to \ell^\pm \widetilde{\nu}^* \to 
\ell^\pm (\ell^+\ell^-)$ and the other as $\widetilde{\chi}_i^\pm \to 
\nu \widetilde{\ell}^{\pm *} \to \nu (\ell^\pm \nu)$. Other processs, 
such as production of a pair of sleptons (each decaying to a lepton and 
missing energy) in association with an $\ell^+\ell^-$ pair, are 
sub-dominant. In all these cases, at least one of the final state 
leptons would tend to be much harder than in the UED case, and the upper 
cut $p_T^\ell \leq 40$~GeV may be enough to virtually eliminate the 
$R$-parity-violating alternatives at the ILC. The $p_T^\ell \leq 
100$~GeV cut at the CLIC would still be a useful one, but probably much 
less efficient in removing this alternative.

\item In a model with large extra dimensions, it is possible to have 
several underlying SM diagrams (e.g. $ZZ$ production) ending in a 
$4\ell$ final state in association with a tower of invisible KK 
gravitons radiated from any leg or vertex, e.g. $e^+e^- \to ZZG_n \to 
(\ell^+\ell^-)(\ell^+\ell^-)\not{\!\!E}_T$. Some of these events will be 
removed by the $p_T^\ell$ cut, and some can be eliminated by vetoing 
events where the $\ell^+\ell^-$ invariant mass peaks at $M_Z$, but there 
would remain a substantial irreducible background which may prove 
difficult to isolate from the UED signal.

\end{enumerate} 
\noindent The above list is illustrative, but not exhaustive. There 
exist possibilities galore, such as production of chargino pairs in 
association with $\ell^+\ell^-$, or processes involving graviton 
resonances, unparticles, and other exotica. As before, the best scenario 
would arise when we have accumulated a large number of events and 
more-or-less eliminated the extra $W'$, $Z'$ and graviton alternatives, 
while the $p_T^\ell$ cut has similarly eliminated the 
$R$-parity-violating version of SUSY. However, the third of the above 
alternatives, viz. heavy neutralino production and decay, is irreducible 
by kinematic means, and must still remain an alternative explanation of 
an excess in $4\ell + \not{\!\!E}_T$ events.

\bigskip\noindent Signals for UED at the ILC (or CLIC) represent, 
therefore, a typical case of the inverse problem, since there are 
several rival models contending for an explanation of each signature. 
The key to solving this must lie in considering more than one signal 
cross-section (or distribution) simultaneously \cite{RaRaC}. Recent 
studies \cite{ArHam} of the inverse problem at the LHC suggest that it 
is convenient to define a `signature space' of different signals, into 
which every point in the parameter space of new physics models will map 
uniquely. The map may not, however, be invertible, since the same 
signature may arise from different kinds of new physics, or even from 
different choices of parameters in the same model. Nevertheless, if 
different models map into different parts of the `signature space', then 
the experimental data will pick up the correct model (or class of 
models) effortlessly. We now demonstrate that such could indeed be the 
case for the UED --- and rival models --- in the light of the signals 
discussed above.

\begin{figure}[h]
\centerline{ \epsfxsize=17.0 cm\epsfysize=6.5cm \epsfbox{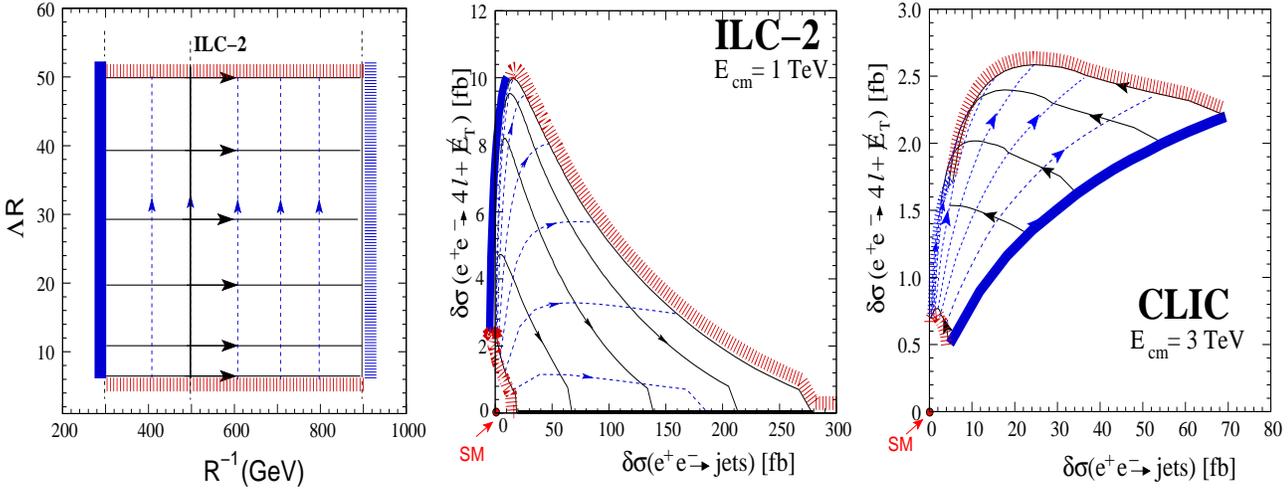} }
\vskip -10pt
\caption{{\footnotesize\it Illustrating the correlation of two different 
cross-sections to obtain a distinctive signature space for UED at the 
ILC-2 and the CLIC. The parameter space is shown in the box on the 
extreme left, while the excess (over SM) cross-sections are shown in the 
central box for the ILC-2 and on the extreme right for the CLIC.  Other 
conventions are explained in the text.}}
\label{fig:Correlation}
\end{figure}

\bigskip\noindent In Figure~\ref{fig:Correlation} we have shown three 
such plots. The plot on the extreme left shows the UED parameter space 
discussed in the text. This may be thought of as a two-dimensional 
section of the multi-dimensional `theory space' at ILC, which 
encompasses all new physics models and all their parameters.  In the 
centre and on the right we show the `signature space' at the ILC-2 and 
CLIC respectively, i.e. once again a two-dimensional section in the 
space of all possible signals, by plotting our predictions for the 
excess contributions (above SM) in the dijet and $4\ell + \not{\!\!E}_T$ 
cross-sections along the two axes in a plane. In order to generate these 
numbers the following kinematic cuts were used:

\noindent $\bullet$ {\it Dijets}: \\
\hspace*{3mm} ($i$)~~~Each `jet' must have transverse momentum $p_T^{\rm 
jet} \geq 100$~GeV. \\
\hspace*{3mm} ($ii$)~~Each `jet' must have pseudorapidity $|\eta_{\rm 
jet}| \leq 2.5$. \\
\hspace*{3mm} ($iii$)~The dijet invariant mass must satisfy $M_{\rm JJ} 
\geq 600$~GeV. \\
\hspace*{3mm} ($iv$)~~The angular separation between the jets must satisfy 
$\Delta R_{\rm JJ} \geq 0.7$. 

\noindent $\bullet$ $4\ell + \not{\!\!E}_T$: \\
\hspace*{3mm} ($i$)~~~Each lepton must have transverse momentum $5~{\rm 
GeV} \leq p_T^\ell \leq 40~{\rm GeV}$ at the ILC-2. \\ 
\hspace*{11mm} At the CLIC, the 
corresponding cuts will be $10~{\rm GeV} \leq p_T^\ell \leq 
100~{\rm GeV}$. \\
\hspace*{3mm} ($ii$)~~Each lepton must have pseudorapidity $|\eta_\ell| 
\leq 2.5$. \\
\hspace*{3mm} ($iii$)~The angular separation between the leptons must 
satisfy $\Delta R_{\rm \ell\ell} \geq 0.2$. \\
\hspace*{3mm} ($iv$)~~The missing transverse energy in the system must 
satisfy $\not{\!\!E}_T \geq 100$~GeV. \\
\hspace*{3mm} ($v$)~~~Dileptons of the same flavour in the final state 
must {\it not} have invariant mass $M_{\ell\ell}$ \\
\hspace*{1.2cm} in the range 80--100~GeV. (This is essentially a veto on 
leptons coming from the \\ \hspace*{1.2cm} decay of a $Z$-boson.)

\bigskip\noindent 
In all three plots in Figure~\ref{fig:Correlation}, the following 
conventions are used:
\begin{itemize}

\item The solid (black) lines represent constant values of $\Lambda R$, 
with $R^{-1}$ increasing along the direction of the arrows. It may be 
noted that the arrow direction is reversed when we go from the central 
box (ILC-2) to the right box (CLIC). The reasons are explained below.

\item The dashed (blue) lines represent constant values of $R^{-1}$, 
with $\Lambda R$ increasing along the direction of the arrows.

\item The thick solid (blue) line represents the experimental constraint 
$R^{-1} ~\gtap ~300$~GeV.

\item The vertical (black) line in the box on the extreme left around 
$R^{-1} = 500$ GeV (which ends in dashes) represents the kinematic reach 
of the ILC-2. In the plot in centre, this lies right on the horizontal 
axis. It is not shown on the CLIC plot, on the right.

\item The thick vertical (blue) hatched line in the left box represents 
the dark matter constraint $R^{-1} ~\ltap ~900$~GeV. This is rendered 
hatched rather than solid to indicate that there is a theoretical bias 
in this constraint. There is no such line in the central box because the 
ILC-2 cannot access this part of the parameter space. On the extreme 
right this bound appears very close to the axis.

\item The other (red) hatched lines represent the limits on $\Lambda R$, 
viz.  $\Lambda R ~\gtap ~5$ and $\Lambda R ~\ltap ~50$. Once again, the 
hatching represents the fact that there is a theoretical bias in these 
constraints -- we may recall that these are essentially ballpark values.

\item In the (signature space) plots in the centre and on the right, the 
solid (red) blob at the origin represents the SM prediction, as marked. 
The size of the blob is a very rough indicator of the error level (at 
1$\sigma$) if the integrated luminosity reaches around 100~fb$^{-1}$. We 
assume this ballpark figure both for the ILC-2 and for the CLIC. This 
tiny blob indicates that the UED model will predict a sizable excess in 
both cross- sections, irrespective of the parameters chosen.

\end{itemize}

\noindent Figure~\ref{fig:Correlation} has several interesting features. 
Let us first consider the central box, i.e. the ILC-2 plot. We note that 
this plot is relevant for only the part of the parameter space where 
$300~{\rm GeV} < R^{-1} < 500~{\rm GeV}$, i.e., which is kinematically 
accessible at the ILC-2. As $R^{-1}$ increases towards the kinematic 
limit of 500~GeV, the excess cross-section for $4\ell + \not{\!\!E}_T$ 
falls quite sharply to the SM value (the origin on the plot) while the 
dijet cross-section keeps growing because the increase in electron flux 
as $x \to 1$ is strong enough to mask phase space effects. Of course, 
once the resonant value is crossed, the dijet cross-section immediately 
shrinks back almost to the origin, i.e. the SM value. This is not 
visually apparent in Figure~\ref{fig:Correlation} because that would 
require a very fine scale on the vertical axis, but we have checked 
numerically that this is indeed what happens. An indicator of this 
phenomenon is provided by the `knee' noticeable in each curve close to 
the horizontal axis, which represents a point where the heavier $Z_2$ 
resonance is pushed out of the kinematic range of the ILC-2, while the 
lighter $\gamma_2$ resonance is still accessible through radiative 
effects.

\bigskip\noindent The box on the extreme right shows a similar plot at 
the CLIC, where the most interesting feature is that some of the arrows 
have reversed sign, indicating a fall in the cross-section as $R^{-1}$ 
increases. This is not really a surprise, however. At the CLIC, running 
at $\sqrt{s} = 3$~TeV, the allowed range $R^{-1} = 0.3$--0.9~TeV, 
indicates, that the electrons and positrons producing a resonance will 
typically carry momentum fractions in the ballpark of 0.1 to 0.3, for 
which Figure~3 tells us that the luminosity function is essentially 
flat. Since the maximum mass of resonances in this range of $R^{-1}$ 
will be less than 2~TeV, and the machine energy is 3~TeV, there will be 
no dramatic effects due to phase space either. Thus, the resonant 
cross-section will essentially echo the behaviour of the propagator, 
which, in the narrow-width approximation can be written as
$$
\frac{1}{(\hat{s} - M_2^2)^2 + M_2^2 \Gamma_2^2} \approx 
\frac{1}{M_2\Gamma_2} \delta\left( \hat{s} - M_2^2\right) \ ,
$$
where $M_2$ stands for the mass of the $\gamma_2$ or the $Z_2$, as the 
case may be. The delta function is, of course responsible for setting 
the value of $x_1x_2$. Noting that $M_2 ~\gtap ~2R^{-1}$ and that 
$\Gamma_2$ increases as $M_2$ increases, it becomes clear that the 
resonant cross-section will {\it fall} as $R^{-1}$ increases. It is this 
effect which is reflected in the direction of the arrows on the solid 
lines in the CLIC plot. We reiterate that at the ILC-2 this effect, 
though present, gets masked by the enormous increase in luminosity as 
$R^{-1}$ increases over the small range permissible at this machine.

\bigskip\noindent A glance at the rough error estimate on the SM points 
indicated in Figure~\ref{fig:Correlation} shows that if the value of 
$R^{-1}$ is kinematically accessible, a clear signal for new physics may 
be obtained at the ILC-2 and CLIC, for a wide range of parameters of the 
UED model. The question one may then ask is whether this signal can be 
uniquely identified as due to a UED. In the earlier part of this section 
we have seen a sample of new physics models which could yield similar 
signals. We now contend that these will have different characteristics 
when mapped on to the signature space illustrated (for ILC-2 and CLIC) 
in Figure~\ref{fig:Correlation}. In order to establish this argument, we 
take up these alternative models one at a time.

\begin{itemize}

\item {\it Extra neutral gauge bosons}: As we have argued above, at the 
ILC-2 we can easily have a $4\ell + \not{\!\!E}_T$ signal from the 
production of a pair of $Z'$ bosons in association with a radiated 
photon $\gamma^\ast$ or a $Z$ boson. This would, obviously, happen at 
the ILC-2 only if $M_{Z'} < \sqrt{s}/2 \simeq 500$~GeV. The $Z'$ could 
also form a resonance\footnote{Or even a pair of resonances, if there is 
a nearly-degenerate $Z''$.} in the $e^+e^- \to JJ$ cross-section, and 
hence appear as a point in the signature space overlapping with the 
portion occupied by the UED signal. However, at the ILC-2, the position 
of the resonance in the invariant mass plot shown in 
Figure~\ref{fig:ILC} would be a dead giveaway, since any $n = 2$ 
resonance in UED will lie approximately at $\sqrt {s} \approx 2R^{-1} 
~\gtap ~600$~GeV, whereas the $Z'$ resonance must perforce lie at 
$\sqrt{s} = M_{Z'} < 500$~GeV.

\bigskip\noindent A more interesting possibility at the ILC-2 is a 
single heavy $Z'$ boson, with $M_{Z'} > 500$~GeV, which not only creates 
a resonance in the dijet cross-section, but is also produced in 
association with a lepton pair and an {\it ordinary} $Z$ boson. For 
example, we can have a standard $e^+e^- \to \ell^+\ell^-$ process, with 
both the $Z$ and the $Z'$ radiated off the leptonic legs. Now, if the 
$Z'$ boson decays invisibly ($Z' \to \nu\bar{\nu}$), and the $Z$ decays 
to a $\ell^+\ell^-$ pair, there may be a significant number of events 
with four soft ($p_T^\ell < 40$~~GeV) leptons. In this case, however, at 
least two of the leptons will have an invariant mass peaking at $M_Z$, 
and this configuration would be eliminated by the cuts used to generate 
Figure~\ref{fig:Correlation}. If, on the other hand, it is the $Z$ boson 
which decays invisibly, and the $Z'$ which gives a dilepton, then there 
will be a peak in the invariant mass of at least one pair of leptons. 
This peak should match the resonant peak in the dijet cross-section, 
thus giving away their common $Z'$ origin. This can be used to 
discriminate from the UED model, where there should be no peaking in 
$M_{\ell\ell}$, and where the dijet resonances are roughly twice as 
heavy as the $Z_1$.  Pinning down a UED model at the ILC-2 may, 
therefore, have to wait until enough statistics is accumulated to make 
an unambiguous statement about the presence or absence of a peak in the 
invariant mass spectrum of all the possible lepton pairings.

\bigskip\noindent At the CLIC, running at $\sqrt{s} = 3$~TeV, a $Z'$ 
pair can be produced so long as $M_{Z'} < \sqrt{s}/2 \simeq 1.5$~TeV. 
This invalidates the argument used above for the ILC-2, since we could 
very well have a $Z'$ in the mass range 600~GeV to 1.5~TeV, which can be 
pair-produced as well as excite a resonance. Moreover, the larger 
bin-size at the CLIC would make it more likely for a {\it single} 
resonant peak to be observed. The signals from a model with a $Z'$ boson 
could, therefore, mimic the UED signal in every particular, except for 
one, viz. the fact already pointed out above: that in the $4\ell + 
\not{\!\!E}_T$, at least one pair of leptons will show an invariant mass 
peak corresponding to the $Z'$ boson mass --- and this will match the 
resonant value of dijet invariant mass. Presumably the higher luminosity 
of the CLIC would make it easy to make an early identification of peaks 
in the dilepton invariant mass as envisaged here.

\item {\it Extra charged gauge bosons}: We have argued earlier, that the 
production of a pair of heavy $W'^\pm$ gauge bosons in association with 
an $\ell^+\ell^-$ pair can closely mimic the $4\ell + \not{\!\!E}_T$ 
signal from UED. However, in this case, no resonance will be seen in the 
dijet invariant mass, simply because the initial and final states are 
charge-neutral. The observed point in the signature space shown in 
Figure~\ref{fig:Correlation} will lie somewhere on the vertical axis, 
and hence, be clearly distinct from a UED signal. This will be as true 
at the CLIC as at the ILC-2.

\bigskip\noindent It is legitimate to ask the question: what will happen 
if there are both $Z'$ resonances as well as $W'$ states which can be 
pair-produced in $e^+e^-$ collisions? Such combinations of states do 
exist in, for example, left-right symmetric models. At the ILC-2, 
existing constraints on such models preclude pair-producion of the extra 
charged gauge bosons, but at the CLIC, this may be a real possibility. 
It is true that in most of the existing models the $W'$ would be 
more-or-less as heavy as the $Z'$ (if not heavier) and not about half as 
massive as is envisaged in this work\footnote{This is not to say that 
one cannot devise a model in which this feature appears --- even if it 
is done only to play the Devil's advocate.}. However, if we confine 
ourselves to a signature space of $4\ell + \not{\!\!E}_T$ signals versus 
dijet signals, any such model will certainly yield nonzero results for 
both.  Fortunately, the leptonic decays of a heavy $W'$ will tend to 
produce hard leptons which would again fall foul of the $p_T^\ell < 
40~(100)$~GeV cut imposed on the $4\ell + \not{\!\!E}_T$ signal at the 
ILC-2 (CLIC). Ultimately, therefore, this veto on hard leptons will 
remove a large part of the cross-section and thus it may be expected 
these models will map to points very close to (but not exactly on) the 
abscissa in the signature space of Figure~\ref{fig:Correlation}.

\item {\it Massive gravitons}: Let us first consider the case of large 
extra dimensions~\cite{ADD}. We have seen that a tower of invisible but 
massive graviton states can closely mimic the UED model for the $4\ell + 
\not{\!\!E}_T$ signal. However, when we consider the $e^+e^- \to JJ$ 
process, there should be an excess cross-section, but {\it no} 
identifiable resonance(s) --- the excess cross-section forming a 
continuum in invariant mass \cite{HLZ}. If the technique adopted to 
identify a resonance is not inspection, but, for example, a $3\sigma$ 
deviation from the SM in one or more bins of $M_{JJ}$, it should be 
designed with a veto on an excess cross-section which spreads across 
several bins. Since, the horizontal axis in Figure~\ref{fig:Correlation} 
is plotted assuming a resonance, we can assume the excess {\it resonant} 
$e^+e^- \to JJ$ cross-section to vanish, so that the large extra 
dimensions model will eventually predict a point lying on the vertical 
axis.

\bigskip\noindent If we consider a single heavy graviton $G_1$, as 
predicted in the Randall-Sundrum model \cite{RS}, a resonance in the 
$e^+e^- \to JJ$ cross-section is predicted, and this will resemble the 
UED case when there are overlapping $\gamma_2, Z_2$ peaks.  Such a 
graviton state can also be produced in association with a $4\ell$ state, 
and decay invisibly as $G_1 \to \nu\bar{\nu}$. Equivalently, it can be 
produced in association with an $\ell^+\ell^-\not{\!\!E}_T$ signal and 
itself decay to another $\ell^+\ell^-$ pair. In either case we obtain a 
$4\ell + \not{\!\!E}_T$ signal. Thus, the presence of a massive graviton 
$G_1$ will produce an overlap with the UED model in the signature space 
of Figure~\ref{fig:Correlation}. There are now two ways to distinguish 
it from the UED signal, both requiring a fairly large number of events. 
A purely kinematic method would be to study the invariant mass 
distribution of every lepton pair in the $4\ell + \not{\!\!E}_T$ final 
state. Obviously, for all the events involving a graviton decay $G_1 \to 
\ell^+\ell^-$, the invariant mass will cluster around $M_{G_1}$, which 
should match with the resonance in $e^+e^- \to JJ$. This matching cannot 
happen in the UED model, where the $Z_1$ bosons leading to missing 
energy and momentum are about half as massive as the $\gamma_2, Z_2$ 
resonances. Another method would be to reconstruct the angular 
distribution $d\sigma/d(\cos\theta^*)$ of the dijet states in the 
centre-of-mass frame and see if they exhibit the characteristics of 
spin-1 particles, i.e. $\propto P_1(\cos\theta^*)$, or spin-2 particles, 
i.e. $\propto P_2(\cos\theta^*)$, where $P_{1,2}$ denote the Legendre 
polynomials of order 1 and 2 respectively. This second method will have 
errors arising from imperfections in jet reconstruction, but will gain 
vastly in the number of events when compared with the first, purely 
kinematic, method. Hence it would probably be the method of choice 
should a signal be seen in the early days of ILC-2, with the other 
method chipping in as a confirmatory test when enough statistics have 
accumulated.
 
\item {\it Supersymmetry with and without $R$-parity conservation}: We 
have argued above that if $R$-parity is conserved, a $4\ell + 
\not{\!\!E}_T$ final state can easily arise in SUSY from the production 
of a pair of heavy neutralinos decaying to the invisible LSP. However, 
the same conservation principle which renders the LSP invisible also 
precludes any of the heavy supersymmetric particles from appearing as 
resonances in $e^+e^- \to JJ$. Hence, the predicted point will lie along 
the vertical axis in the signature space of 
Figure~\ref{fig:Correlation}.

\bigskip\noindent The distinction becomes somewhat more difficult if 
$R$-parity is violated, for then we can have sneutrino resonances in the 
$e^+e^- \to JJ$ cross-section as well as an excess in the $4\ell + 
\not{\!\!E}_T$ signal due to a pair of decaying neutralino LSPs. Here, 
as in the case of a massive graviton resonance, we can use a purely 
kinematic method and/or a dynamical method relying on the fact that the 
sneutrino is a scalar. The dynamical method is quite simple, since it 
involves the same construction of $d\sigma/d(\cos\theta^*)$ for the 
dijet final state as described above. If the exchanged particle is a 
scalar, this should be $\propto P_0(\cos\theta^*)$, i.e. constant (apart 
from kinematic effects near $\cos\theta = \pm 1$ which are due to the 
acceptance cuts). Any substantial variation in $d\sigma/d(\cos\theta^*)$ 
should immediately rule out the sneutrino hypothesis. At the same time, 
we may note that the kinematic cut $p_T^\ell \leq 40$~GeV can also 
remove most of the $4\ell + \not{\!\!E}_T$ signal at the ILC for an LSP 
mass of around 100~GeV or more. The corresponding cut, $p_T^\ell \leq 
100$~GeV at the CLIC will be less efficient, but would still reduce the 
LSP decay signal considerably. Assuming this to be the case, the 
predicted point in the signature space will lie practically on the 
horizontal axis. At the ILC-2, this will be similar to the UED 
prediction with $R^{-1} ~\ltap ~500$~GeV, but, of course, when we go to 
the CLIC, the UED prediction will no longer be on the horizontal axis 
(as shown in the right-most box in Figure~6), whereas that from 
$R$-parity-violating SUSY will continue to be so. One or other of the 
two methods will suffice to pick out UED from $R$-parity-violating SUSY.

\end{itemize} 

\noindent The above list of models is illustrative, but by no means
exhaustive\footnote{There are already many other suggested models in the
literature, such as little Higgs models with $T$-parity, unparticles,
etc., which could also, in principle, have been included in the
argument. Indeed, it would be foolhardy to try and make an exhaustive
list, for that would not only be underestimating human ingenuity, but
also be making the hubristic assumption that Nature cannot have any more
surprises for us.}. It may happen that the reality at the TeV scale
comprises not just one of the above models, but a combination of two or
more of these. For example, if we have a supersymmetric model with an
extended gauge sector, i.e. an extra $Z'$ boson, we may expect an excess
in the $4\ell + \not{\!\!E}_T$ signal due to the production of a slepton
pair in association with a $Z$ or a $Z'$ boson which then decays to
leptons. At the same time, the $Z'$ could give rise to resonances in the
dijet cross-section. However, some of the tests suggested above could
still be used to distinguish this scenario from the UED case. For
example, the leptons coming from the $Z'$ decay would not only tend to
be hard, unlike those coming from a $Z_1$ to $\ell_1$ to $\gamma_1$
cascade decay, but would also exhibit an invariant mass peak at the $Z'$
mass which would match with the resonant mass of the dijets. Thus, the
arguments presented above serve to illustrate the fact that it is indeed
very difficult to create a model which would not have {\it some}
kinematic or dynamical difference from the UED model so far as its
signals at the ILC-2 go, especially when we combine two or more signals,
and add on a spin measurement. It follows that if UEDs are indeed the
new physics to be discovered at the next generation of accelerators,
then it should be possible to make a unique statement to that effect,
using some very simple kinematic constructions and physical arguments.

\vskip 10pt
\section{Summary and Conclusions}

\noindent We have already stressed in the Introduction that the LHC does 
not lend itself readily to UED searches. It is likely that UED searches 
at the LHC would require considerable ingenuity and the collection of a 
large amount of data before any definitive conclusion can be reached. 
The strategies discussed in this paper have, therefore, deliberately 
been kept independent of the possible discovery of UED-s or some other 
form of new physics at the LHC --- partly because it would be mere 
speculation to talk of a post-LHC scenario at the present juncture, and 
partly because we feel that the ILC (or CLIC) should be considered as an 
independent experiment in its own right.

\bigskip\noindent In this work, we have focussed on the efficacy of 
radiation from the colliding beams in an $e^+e^-$ collider in producing 
an energy spread and shown that this leads to enhanced cross-sections 
for resonant processes involving $n=2$ KK excitations of the photon and 
the $Z$-boson in a model with universal extra dimensions. For a few 
benchmark points, we have shown that one can get sharply-defined 
resonances in the invariant mass of final-state dijets. Focussing on 
these, and combining with the pair production of $Z_1$ resonances 
(followed by leptonic decays of the $Z_1$-s), we have shown how to 
create a two-dimensional `signature space' and use it to distinguish UED 
signals from that of other, competing models for new physics. As we have 
seen, there is no `smoking gun' signal which can be uniquely identified 
as due to UEDs, but the simple two-dimensional signature space of 
Figure~\ref{fig:Correlation}, with at most, a spin measurement added, 
would be more-or-less sufficient to identify an underlying UED model 
uniquely.

\bigskip\noindent A clear UED signal, corresponding to a point in the 
signature space well away from any of the axes would immediately tell us 
the values of the parameters $R^{-1}$ and $\Lambda R$, as corresponding 
to a unique point in the grid of Figure~\ref{fig:Correlation}. It is, 
indeed, quite a remarkable feature of the UED model that almost complete 
disambiguation can be achieved with so simple an analysis, especially 
when we compare it with the intricate and computation-intensive methods 
which have been employed to study SUSY and string-inspired models 
\cite{ArHam}. While our analysis admittedly lacks sophistication in the 
reconstruction of jets, we contend that for a fairly high number of 
events, the predictions of a more elaborate simulation at an $e^+e^-$ 
collider would closely resemble the ones presented in this paper. 
Moreover, we have chosen two final states, viz., dijets and $4\ell$ plus 
missing $p_T$, for which data will certainly be collected and stored at 
any high energy $e^+e^-$ machine. The search for and identification of 
UED-s can, therefore, be carried out without any extra cost, as it were, 
in data storage or even manpower, using the techniques described in this 
paper. Our analysis thus has the twin virtues of economy and simplicity. 
On that positive note, we end the discussion.
     
\bigskip\noindent {\small {\sl Acknowledgments}: The authors would like 
to thank the organisers of LCWS-06 (Bangalore, 2006) and WHEPP-X 
(Chennai, 2008) where some of this work was done. We acknowledge partial 
financial support from the University Grants Commission, India (BB), the 
Department of Science and Technology, Government of India (AK), the 
Board of Research in Nuclear Science, Government of India (AK) and, 
finally, the Academy of Finland (SKR).}

\end{document}